\documentclass[prb,aps,twocolumn,superscriptaddress,10pt,showpacs,longbibliography]{revtex4-1}
\usepackage{graphicx}
\usepackage{epstopdf}
\usepackage{color}
\usepackage{amsmath,amssymb}
\usepackage{fixmath}

\newcommand{\angstrom}{\mbox{\normalfont\AA}}
\usepackage [english]{babel}
\usepackage [autostyle, english = american]{csquotes}
\MakeOuterQuote{"}
\usepackage{textgreek}
\usepackage[export]{adjustbox}
\usepackage{siunitx}
\usepackage{xcolor}
\usepackage{hyperref}
\begin{document}

\title{Anharmonicity and scissoring modes in the negative thermal expansion materials ScF$_{3}$ and CaZrF$_{6}$}

\author{T. A. Bird} 
\affiliation{Department of Chemistry, University of Warwick, Gibbet Hill, Coventry, CV4 7AL,United Kingdom}

\author{J. Woodland-Scott}
\affiliation{Department of Chemistry, University of Oxford, Inorganic Chemistry Laboratory, South Parks Road, OX1 3QR Oxford, UK}

\author{L. Hu}
\affiliation{Department of Physical Chemistry, University of Science and Technology Beijing, Beijing 100083, China}

\author{M. T. Wharmby}
\affiliation{Deutsches Elektronen-Synchrotron (DESY), Notkestr. 85, 22607 Hamburg, Germany}

\author{J. Chen}
\affiliation{Department of Physical Chemistry, University of Science and Technology Beijing, Beijing 100083, China}

\author{A. L.  Goodwin}
\affiliation{Department of Chemistry, University of Oxford, Inorganic Chemistry Laboratory, South Parks Road, OX1 3QR Oxford, UK}

\author{M. S. Senn}
\email{m.senn@warwick.ac.uk}
\affiliation{Department of Chemistry, University of Warwick, Gibbet Hill, Coventry, CV4 7AL,United Kingdom}

\date{\today}

\begin{abstract}
	We use a symmetry-motivated approach to analysing X-ray pair distribution functions to study the mechanism of negative thermal expansion in two ReO\textsubscript{3}-like compounds; ScF\textsubscript{3} and CaZrF\textsubscript{6}. Both average and local structure suggest that it is the flexibility of M-F-M linkages (M = Ca, Zr, Sc) due to dynamic rigid and semi-rigid "scissoring" modes that facilitates the observed NTE behaviour. The amplitudes of these dynamic distortions are greater for CaZrF$_6$ than for ScF$_3$, which corresponds well with the larger magnitude of the thermal expansion reported in the literature for the former. We show that this flexbility is enhanced in CaZrF$_6$ due to the rock-salt ordering mixing the characters of two of these scissoring modes. Additionally, we show that in ScF$_3$ anharmonic coupling between the modes responsible for the structural flexibility and the rigid unit modes contributes to the unusually high NTE behaviour in this material.
\end{abstract}

\maketitle

\section{Introduction}

Research into materials that contract upon heating, termed negative thermal expansion (NTE) materials, has been steadily increasing over the past 30 years. The significance of the phenomena was first underlined by Sleight \textit{et al.} in 1996\cite{Evans1996} by linking the large, isotropic NTE of ZrW$_{2}$O$_{8}$ to the crystal structure of the material, opening up the field to synthesis of new compounds. Since then, this field has been expanded to a wider range of materials, including simple oxides (such as Cu$_{2}$O\cite{Dapiaggi2003} and ReO$_{3}$\cite{Chatterji2008,Chatterji2009}) and metal-organic frameworks\cite{Grobler2013,Lock2010}.\\
The rigid unit mode (RUM) model is a common way to explain the origin of NTE\cite{Dove2016}. Materials made from rigid polyhedra have a significant energy barrier to distortions of the polyhedra, but a low barrier to collective dynamics such as rotations. These modes are often low in energy so have a significant contribution to the coefficient of thermal expansion, and they can lead to NTE via the tension effect - if two linked bonds are straight, or nearly straight, and stretching the bonds would take a large amount of energy, a transverse displacement of the central atom would pull the two other atoms closer together, resulting in a local decrease in volume, the magnitude of which would increase when the temperature is raised\cite{Barrera2005}. ReO$_{3}$, a material made from corner-sharing ReO$_{6}$ octahedra (hence can be thought of as an A-site deficient perovskite), is commonly used to illustrate this model due to the complexity of the motion in more typical NTE materials such as ZrW$_{2}$O$_{8}$. The octahedra in this material are expected to dynamically rotate in an out-of-phase manner with respect to their neighbouring units about their average positions, resulting in a contraction of the structure whilst the material remains, on average, cubic\cite{Chatterji2008}. It is two compounds similar to ReO$_{3}$ that are studied herein; the isostructural ScF$_{3}$, and the A-site deficient double perovskite CaZrF$_{6}$. Metal trifluorides adopting the ReO$_3$ structure typically undergo a transition from the $Pm\bar{3}m$ cubic structure to a rhombohedral phase ($R\bar{3}c$) upon cooling, via long range ordering of the MF$_{6}$ octahedra (a\textsuperscript{-}a\textsuperscript{-}a\textsuperscript{-} in Glazer notation). The dynamic motion of these tilts was expected to be the mechanism for NTE in ScF$_{3}$\cite{Greve2010} supported by the fact that a phase transition to the rhombohedral tilt phase is observed under hydrostatic pressure of $~0.7$ GPa at ambient temperature\cite{Greve2010,Handunkanda2015} and in the related material CoZrF$_6$, whose high temperature phase is isostructural to CaZrF$_6$\cite{Hancock2015}. NTE is observed at a range of temperatures above the phase transition, but below it, once the phonon mode associated with the RUM has been “frozen in”, strong PTE is observed. Previous studies of these materials have shown large displacements of the fluoride ions perpendicular to the M-F-M bonds (M = Sc, Ca, Zr)\cite{Hu2016,Gupta2018}, consistent with a polyhedral rocking mechanism for NTE. Other studies have challenged the RUM model, concluding that only certain bonds were rigid\cite{Tucker2005,Sanson2016}, rather than entire polyhedra, and that bond bending could be a contributor to NTE\cite{Sennova2007,Senyshyn2010}.\\

Several studies have been performed recently to try and ascertain the origin of NTE in these materials. X-ray pair distribution function (PDF) analysis of two materials in the cubic MZrF$_6$ (M = Ca, Ni) series has shown that differing degrees of flexibility in M-F linkages results in isostructural materials having very different thermal expansion properties\cite{Hu2016a}. Lattice dynamics calculations of ScF$_3$ performed by Li \textit{et al.}\cite{Keith2011} showed mostly soft lattice modes that distorted the ScF$_6$ octahedra, however a 3 x 3 x 3 grid of unit cells was chosen, which excludes the zone-boundary wavevectors which the RUMs are confined to. Molecular dynamics simulations on the general ReO$_3$ structure\cite{Schick2016}, with variable interaction strengths, suggest a degree of flexibility in the octahedra enhances NTE. Another conclusion from these simulations was that a weaker anion-anion nearest neighbour interaction enhances NTE, which is supported experimentally by the greater magnitude of NTE in ScF$_3$ compared to ReO$_3$. There is experimental evidence from Raman spectra and inelastic neutron scattering that the large NTE in these materials cannot be accurately predicted with the quasi-harmonic approximation\cite{Sanson2016,Keith2011}, so subsequently lattice dynamics calculations have been done to elucidate the connection between NTE and phonon anharmoncity, since the relatively simple structure compared to other NTE materials allows for a more detailed analysis. These calculations show that cubic\cite{Oba2018} and quartic\cite{Keith2011,Gupta2018} anharmonicity contribute significantly to the temperature dependence of the thermal expansion coefficient. Other simulations have shown that modes with quartic potential can have an enhanced NTE compared to a single-well potential\cite{Sanson2018}.\\
ABO$_3$ perovskites exhibit a wide range of octahedral tilt phase transitions, as classified by Glazer\cite{Glazer1972}, yet do not generally display phonon driven NTE. However, we have recently demonstrated how, by using a symmetry motivated approach to analysing PDF data, we can gain extra information on disorder and dynamics\cite{Senn2016}. Our study on BaTiO$_3$ showed that this method is very sensitive to soft phonon modes of RUM-like character. Here, we use this method to probe the character of the low-lying thermal excitations in the titled compounds, where the amplitudes of such vibrations are believed to be very large.
\section{Experimental Details and Data Analysis}

Scandium trifluoride was used as supplied by Strem Chemicals. Synchrotron radiation X-ray total scattering experiments were conducted at the synchrotron facility PETRA III (beamline P02.1\cite{Dippel2015}) at DESY, Hamburg. A wavelength $\lambda = \SI{0.2070}{\angstrom} $ was used to collect data. Data were collected at temperatures of 125, 140, 147, 152 K and at intervals of 25 K from 175 to 450 K. The obtained 2D images were masked and radially integrated using the DAWN\cite{Basham2015} software. G(r) and D(r) functions were computed using GudrunX\cite{McLain2012} using a $Q_{max} = \SI{21}{\angstrom}^{-1}$. GudrunX was also used to perform background subtraction and sample absorption corrections. 

The CaZrF$_6$ was that prepared \textit{via} a standard solid state synthesis methods in Ref 18. The total scattering data was collected at 11-1D-C APS, Argonne National Laboratory, using a wavelength $\lambda = \SI{0.11798}{\angstrom} $ between 25 and 400 K. The PDFs were computed using PDFGetX2\cite{Qiu2004}, which was also used for background subtraction and sample absorption corrections. A $Q_{max} = \SI{28}{\angstrom}^{-1}$ was used for the analysis presented below.

\subsection{Pair Distribution Function Analysis}
\begin{figure}[t!]
	\hspace*{-1.2cm}
	\includegraphics[width=8.9cm,left]{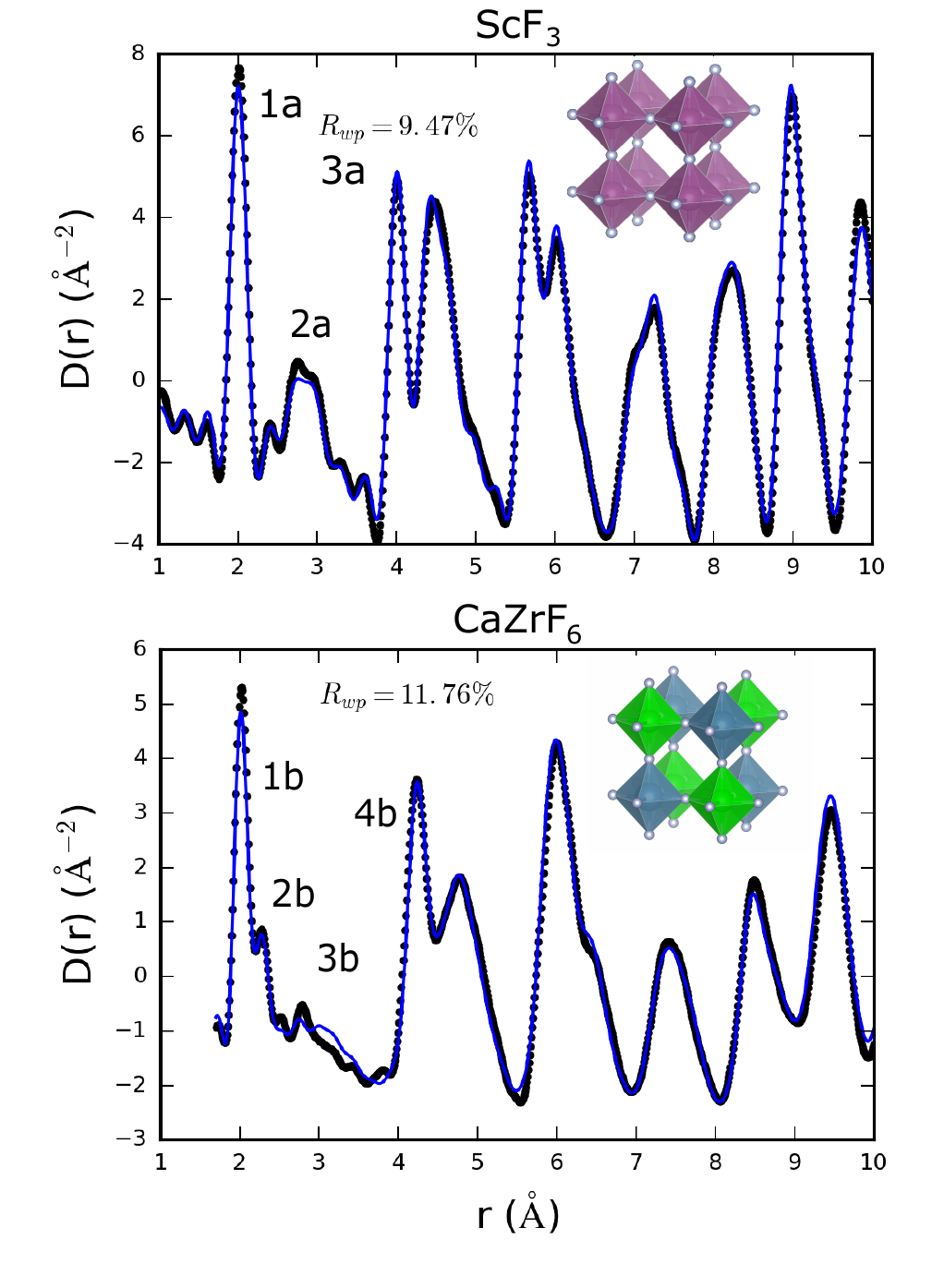}
	\caption{\label{f1} Pair Distribution Function for ScF$_3$ (top) and CaZrF$_6$ (bottom) at 400 K (black circles). A small box fit with the modes belonging to X$_{5}^{+}$ refined is shown for both compounds (blue lines), with the R$_{2}^{-}$ mode additionally refined for CaZrF$_6$. Labelled peaks correspond to Sc-F (1a), F-F (2a, 3b), Sc-Sc (3a), Zr-F (1b), Ca-F (2b) and Ca-Zr (4b).}
\end{figure}

\begin{figure*}[t!]
	\centering
	\includegraphics[width=\textwidth]{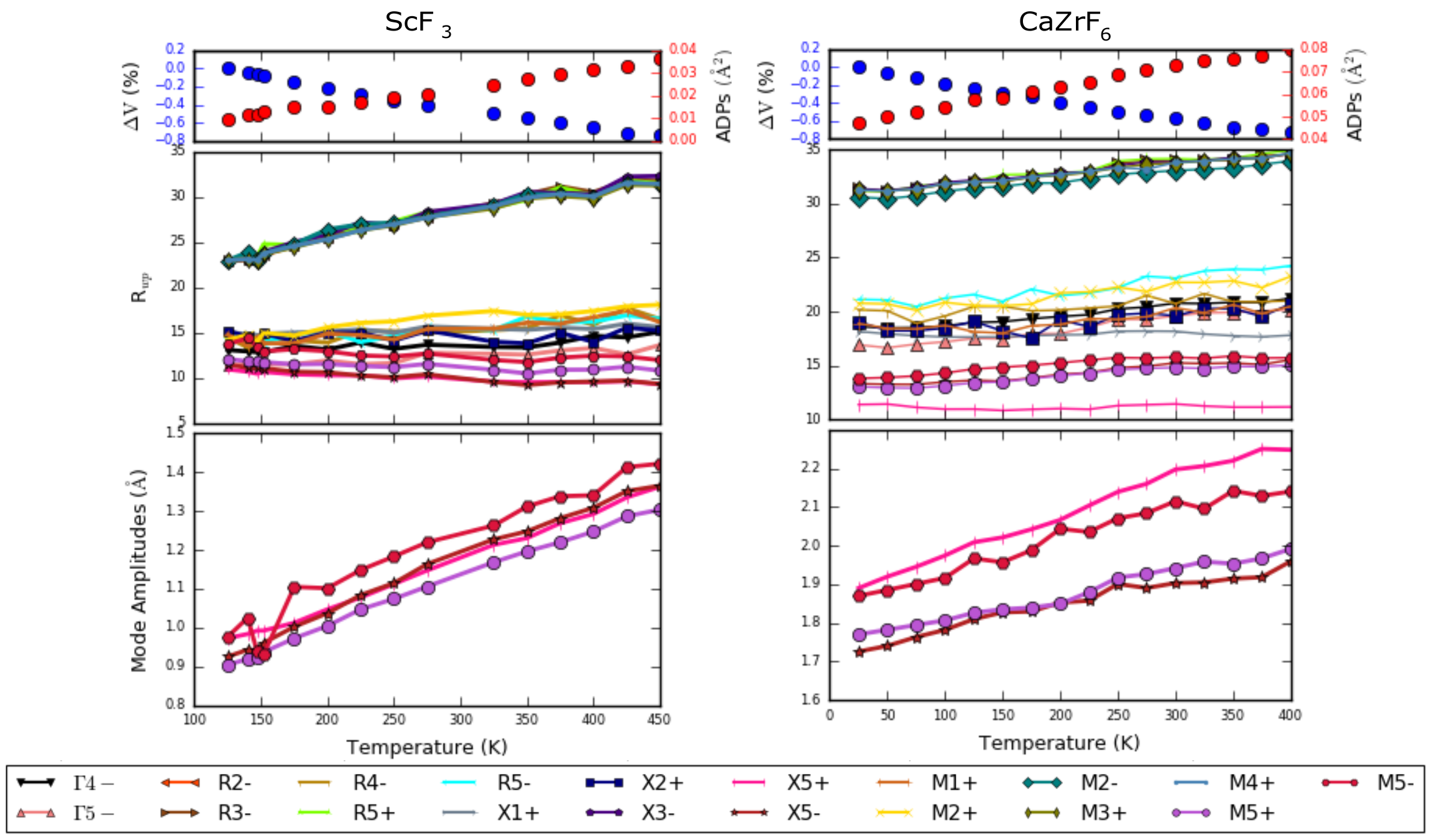}
	\caption{\label{f2} Transverse atomic displacement parameters from Rietveld refinement (top), the best weighted-phase R-factor for each irrep at each temperature (middle) and the boltzmann weighted mode amplitude (bottom). Results for ScF$_3$ are displayed on the left, CaZrF$_6$ on the right.}
\end{figure*}

Some form of modelling is usually required to extract information of interest, such as local distortions of atoms away from their high symmetry positions, from pair distribution functions. The method presented here involves expanding the possible degrees of freedom in terms of symmetry adapted displacements of the zone centre and zone boundary irreducible representations (irreps) of the $Pm\bar{3}m$ A-site deficient perovskite structure. For this analysis we use a parent $Pm\bar{3}m$ perovskite with the A-site at the origin. Symmetry-breaking displacements transforming as the same irrep can be further decomposed into symmetry adapted distortion modes by choosing a sensible basis that reflects the chemistry and crystallographic axes of the structure. The distortion modes have a 1:1 correspondence with phonon eigenvectors in the limit that only one set of atomic displacements transforms as the corresponding irrep. In cases where distortions from different Wyckoff sites transform as the same irrep, the character of the low lying excitations can still be ascertained through refining the relative amplitudes of the individual distortion modes. An overview of the displacements that enter into each irrep is tabulated in a recent paper by Popuri \textit{et al.}\cite{Popuri2019}. For both compounds, ISODISTORT\cite{Campbell2006} was used to generate a model parameterised in terms of symmetry adapted displacements. A 2 x 2 x 2 P1 supercell was used for ScF\textsubscript{3}, since this allows phonon modes with propogation vectors k = [0 0 0], [1/2 0 0], [1/2 1/2 0] and [1/2 1/2 1/2] to be modelled. Whilst this is only a small fraction of possible wave vectors, these are both the ones that PDF data has the greatest sensitivity to and for which our symmetry motivated approach provides the greatest number of constraints. Furthermore, even if the exact wave vectors of the NTE driving phonons are of a longer wavelength, we still expect the character of those phonons to be reflected in our results which probe a shorter wavelength. To generate the parameterisation of CaZrF\textsubscript{6}, a 2 x 2 x 2 supercell of disordered Ca$_{0.5}$Zr$_{0.5}$F$_3$ was used. The cations were then set to be fully ordered to generate the rock-salt ordered structure. In all refinements, the breathing mode about the Ca/Zr (transforming as R$_{2}^{-}$) was refined, making this description equivalent to the published $Fm\bar{3}m$ structure\cite{Hancock2015}. The generated mode listings were output from ISODISTORT in .cif format and then converted to the .inp format of the TOPAS Academic software v6\cite{Coelho2015}. Modes transforming as the same irreducible representation (irrep) were tested simultaneously. An example of the best single-irrep refinement for each compound using this method is shown in Fig. \ref{f1}.
The results shown below (Fig. \ref{f2}) were performed with a fitting range of 1 (ScF$_3$) or 1.7 (CaZrF$_6$) to $\SI{10}{\angstrom}$. The refinements were also done out to a higher radius, however the results were broadly similar for these larger fitting ranges. A comparison between the results for 10 and  $\SI{30}{\angstrom}$ can be seen in the SI (Fig. S3).

The thermal parameters for each site were modelled with a simple quadratic, i.e $b_i = b_{i,low} + ur + vr^2$ where $u$ and $v$ are constant across all sites for each refinement, and $b_{i,low}$ is element dependent. Whilst this does not capture the true physical behaviour of the system, it was found to produce more robust fits to the data (more stable and fewer false minima) than other functional forms of $b_i$, with the results still being consistent with our analysis performed using different functional forms of $b_i$ (see Fig. S2 in the SI). \\
To get an unbiased view of how each irrep influences the local structure, the refinement for each irrep was initiated from randomised starting values of the relevant mode amplitudes. When a minimum was reached, the refined parameters were stored, re-randomised, and a new cycle was initiated. This process was repeated until 25000 iterations were reached (between 300 and 4000 refinements); this process was used to ensure a global minimum was reached for each mode. For refinements of atomic displacements transforming as the \textGamma\textsubscript{4}\textsuperscript{-} irrep, corresponding to ferroelectric type distortions, the amplitudes of modes affecting the metal cations were used to fix the origin; otherwise the mode amplitudes of this irrep would appear artificially high, due to the floating origin of the unit cell. Finally, we note that if the refined mode amplitudes are treated as the mean absolute value of displacement of an harmonic oscillator, then the amplitude of the harmonic motion will be a factor of $\sqrt{2}$ larger than the refined values.
\begin{figure}[t]
	\includegraphics{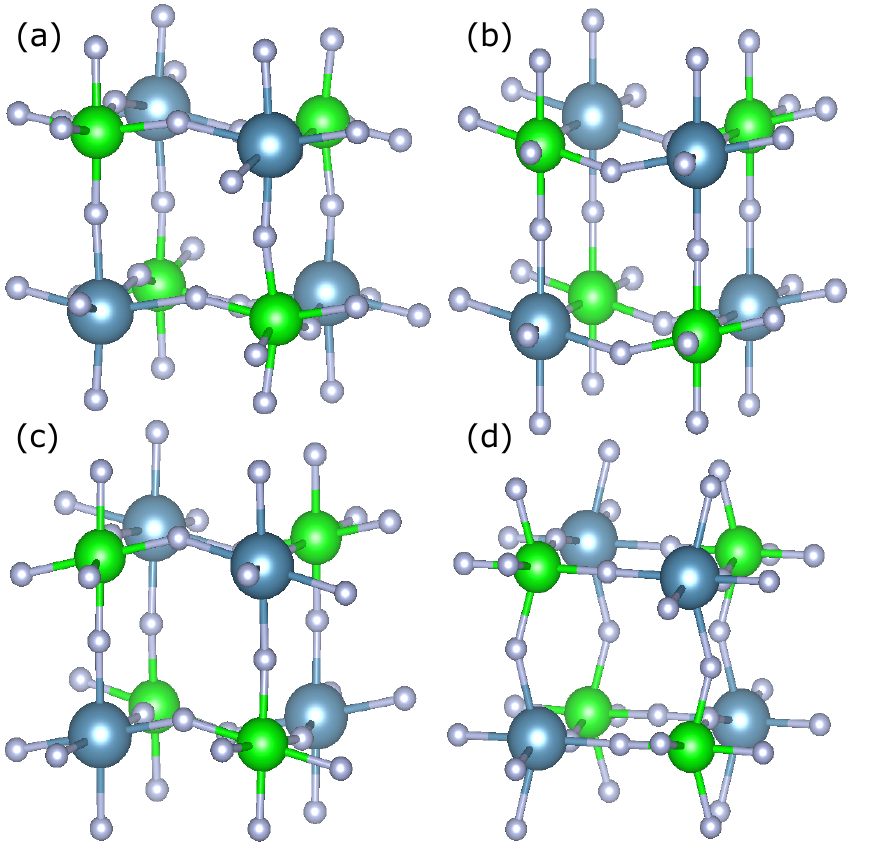}
	\caption{\label{f7} Representations showing the effect of (a) X$_{5}^{+}$, (b) X$_{5}^{-}$, (c) M$_{5}^{+}$ and (d) M$_{5}^{-}$ on the crystal structure of CaZrF6$_6$. The distortions are taken from the refinements at 400 K with the lowest R$_{wp}$ and plotted using the VESTA software\cite{Momma2011}  }
\end{figure}
\begin{figure}[t!]
	\includegraphics{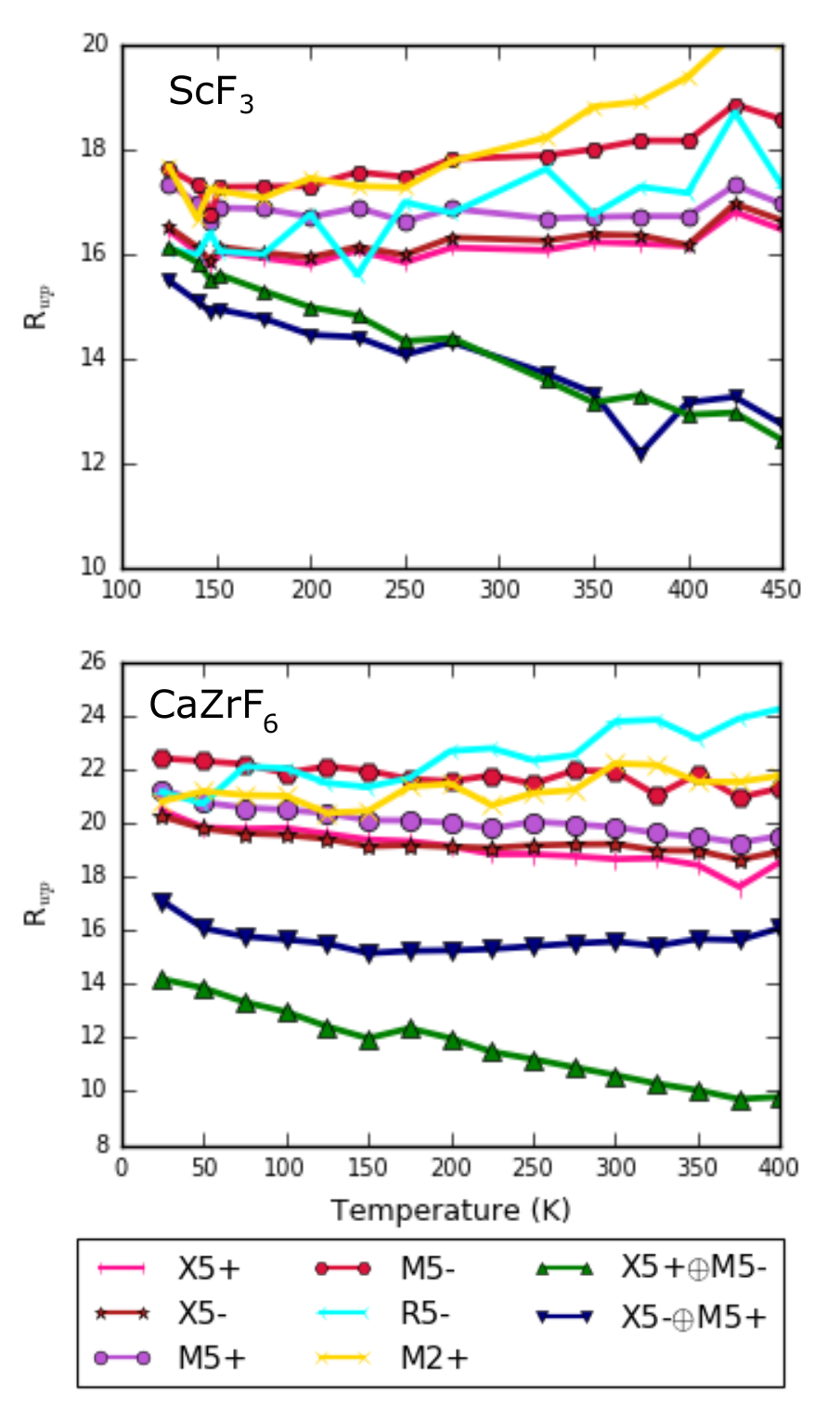}
	\caption{\label{f3} Comparison of weighted R-factors for restricted irreps X$_{5}^{+}$, X$_{5}^{-}$, M$_{5}^{+}$ and M$_{5}^{-}$, unrestricted irreps M$_{2}^{+}$ and R$_{5}^{-}$ and coupled X$_{5}^{+}$$\oplus$M$_{5}^{-}$ and X$_{5}^{-}$$\oplus$M$_{5}^{+}$.}
\end{figure}
\subsection{Constrained Order Parameter Directions}
Some order parameters can have many degrees of freedom associated with them. The exact number is a function of the degeneracy of the propogation vectors, the dimensionality of the irrep and the number of distortions transforming as the irrep. All of these degrees of freedom are described by the collection of symmetry adapted displacements or "distortion modes" that can be labelled accordingly. For example, in the parent structure ($Pm\bar{3}m$) of ScF$_3$ there are 3 types of distortion that transform as X$_{5}^{+}$ which is two dimensional and associated with the triply degenrate k-vector [1/2 0 0], which results in a total of 18 parameters, compared to just 3 for M$_{2}^{+}$ (a triply degenerate single dimensional k-vector) and R$_{5}^{-}$ (a non-degenerate k-vector with 3 dimensions). In our refinements, to facilitate a fairer comparison between irreps, the order parameter direction (OPD) associated with the 3 wave vectors for each distortion have been set to the same values, i.e the general OPD (a,b;c,d;e,f) has been set to (a,b;a,b;a,b). Different distortion modes of the same type associated with the a and b branches of the OPD are allowed to have different values. However to further reduce the degrees of freedom that ratio between a and b across all distortion types that transform as a single irrep are fixed to be constant across different temperature ranges. This reduces the number of parameters for X$_{5}^{+}$ from 18 to 4. Physically, these approximations correspond to a harmonic approximation in which the order parameter directions with respect to the propogation vectors and irrep dimensionality are strictly degenerate in energy. An example of this implementation is given in the SI.

\begin{figure*}[ht]
	\hspace*{-1.2cm}
	\includegraphics{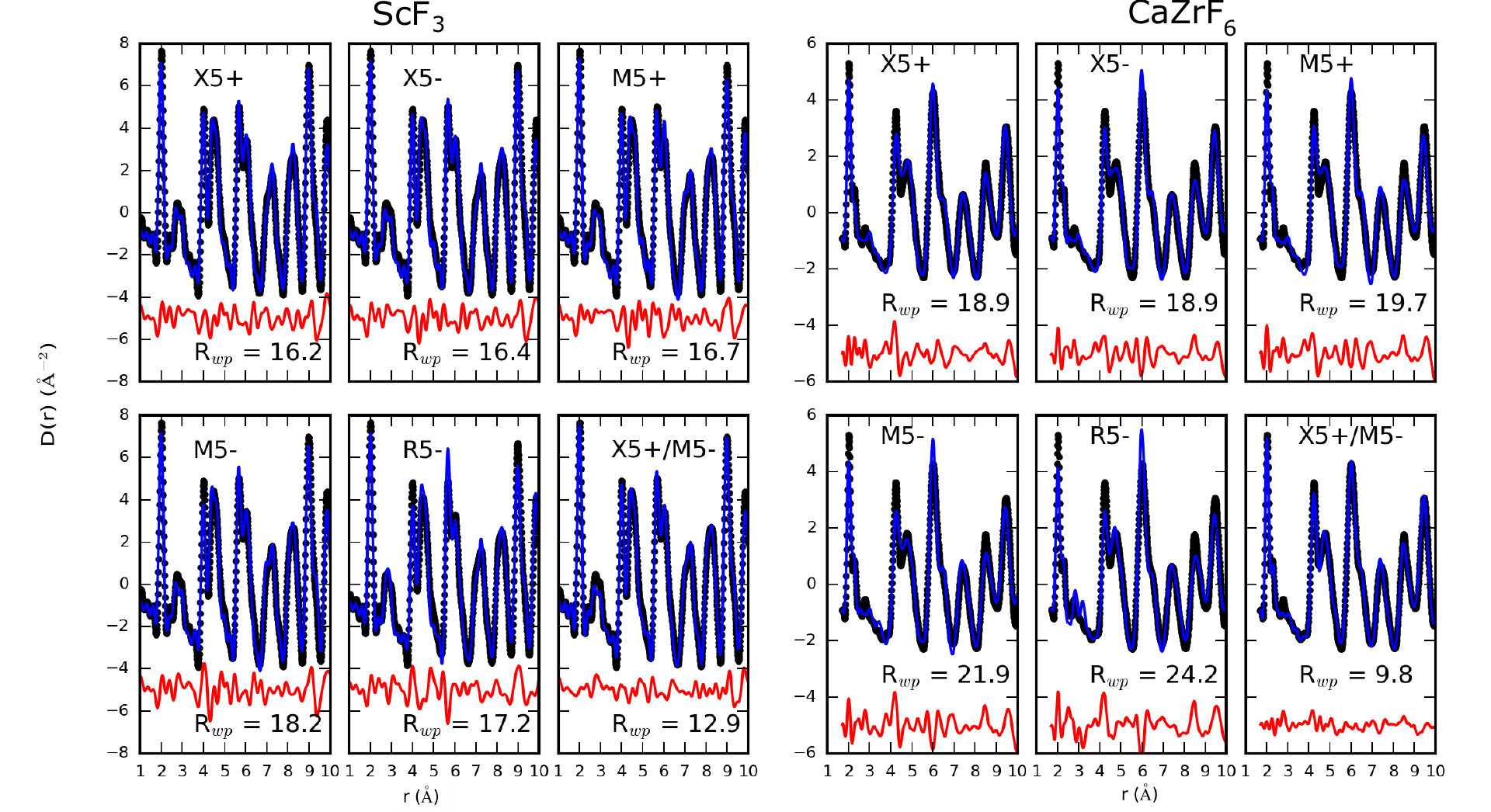}
	\caption{\label{f4} Comparison of fits to ScF$_3$ (left) and CaZrF$_6$ (right) PDF  data at 400 K using restricted X$_{5}^{+}$, X$_{5}^{-}$, M$_{5}^{+}$ and M$_{5}^{-}$, unrestricted R$_5^{-}$ and restricted X$_{5}^{+}$$\oplus$M$_{5}^{-}$}.
\end{figure*}

\section{Results and Discussion}

Rietveld refinement of ScF\textsubscript{3} and CaZrF\textsubscript{6} powder patterns can be used to gain some insight into the NTE behaviour, but can also be misleading; the average structure of both compounds remains cubic over the temperature ranges used, however, this structure fits the pair distribution function quite poorly, with PDFGUI\cite{Farrow2007} refinements of both structures from $1 - \SI{10}{\angstrom}$ having R$_{wp}$s $\approx 18 $ and $20 \% $ for ScF$_3$ and CaZrF$_6$ respectively (see SI Fig S1). The average linear coefficient of thermal expansion (CTE) $ \approx -7.5 \:  ppm \; K^{-1}$ for ScF\textsubscript{3} matches the literature reports well\cite{Greve2010}. The measured CaZrF$_6$ linear CTE, as reported by Hu \textit{et al.} from the same data\cite{Hu2016b}, is $-6.69 \: ppm \; K^{-1}$. In the literature, CaZrF$_6$ is reported to have a magnitude of NTE approximately two to three times that of ScF$_3$\cite{Greve2010,Hancock2015} for the temperature range 25 - 400 K, whereas in these measurements they have quite similar values. The difference to literature reports are in part due to the differing temperature ranges over which CTEs are reported, but may also be due to different strain, morphology and thermal histories of samples\cite{Yang2016,Prisco2013}. The refined atomic displacement parameters (Fig \ref{f2}, top) reveal that most thermal motion of the F ions is perpendicular to the M-F-M linkages (M = Sc, Ca, Zr), indicating that a tensioning of these linkages could be responsible for the observed NTE.\\

Some information can be gained from the PDFs without any modelling. Firstly, the effect of the rock-salt ordering of Ca$^{2+}$ and Zr$^{4+}$ in CaZrF$_{6}$ can be seen in the presence of 2 peaks at  $r \approx \SI{2}{\angstrom}$, compared to just one in ScF$_3$; the greater positive charge of Zr$^{4+}$ compared to Ca$^{2+}$ means the F\textsuperscript{-} ions do not sit at the midpoint of Ca-F-Zr bonds (Fig. \ref{f1}). Secondly, the relative magnitudes of the shorter inter-atomic separations (Sc-F, Sc-Sc; Ca-F, Zr-F and Ca-Zr) means that the magnitude of the mean M-F-M angle (M = Sc, Ca, Zr) must deviate from 180$^{\circ}$. The magnitude of this deviation is larger for CaZrF$_6$ than for ScF$_3$ (see Fig S3). The first peak for ScF$_3$ and the first two for CaZrF$_6$ are noticably less broad than the other peaks, indicating that the M-F bonds are relatively stiff. In contrast, the broadness of the F-F peaks at \textit{circa} $ \SI{3}{\angstrom}$ indicate a propensity for bending of the bond angles within the MF$_6$ octahedra. Little further information can be gained from a simple inspection of the PDFs, hence analysis of the structures has been performed in terms of symmetry-adapted displacements, as described in section II.A.\\

\begin{figure}[t!]
	\hspace*{-1.0cm}
	\includegraphics{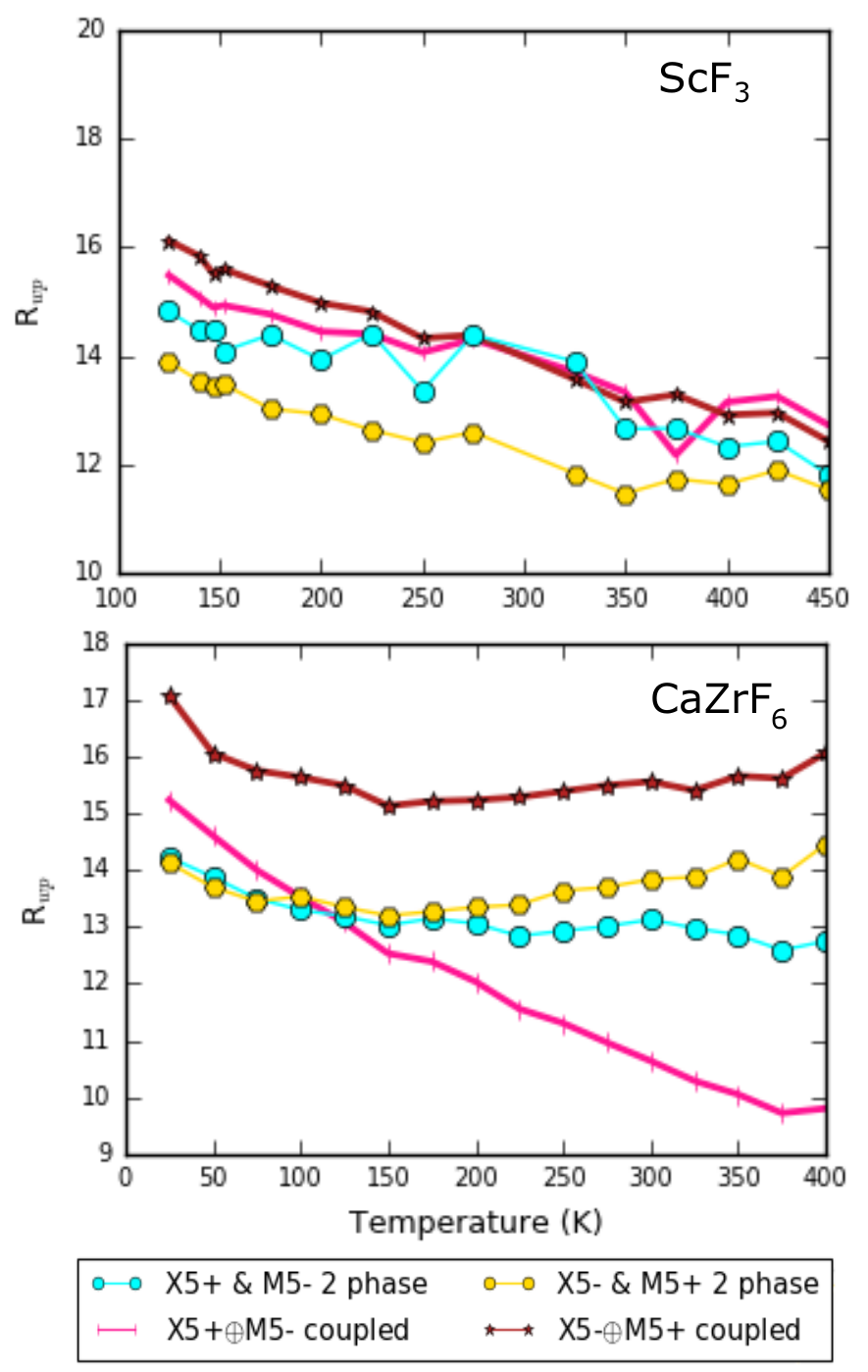}
	\caption{\label{f5} Comparison of coupled and 2-phase fits to PDF data as a function of temperature for ScF$_3$ (top) and CaZrF$_6$ (bottom), as described in the text.}
\end{figure}

\begin{figure}[t!]
	\hspace*{-1.0cm}
	\includegraphics{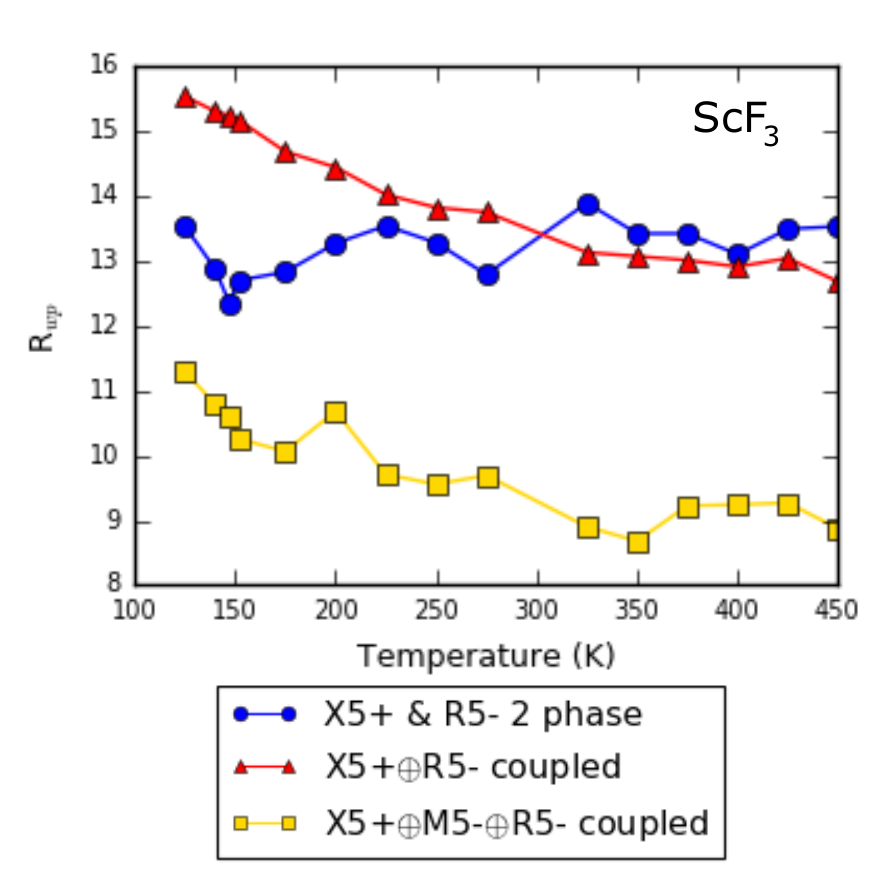}
	\caption{\label{f6} Comparison of fits for X$_{5}^{+}$$\oplus$R$_{5}^{-}$ using a 2-phase model (blue) and a coupled model (red) and X$_{5}^{+}(a,b;0,0;0,0)$$\oplus$M$_{5}^{-}(0,0;c,d;0,0)$$\oplus$R$_{5}^{-}(e,f,g)$ (yellow) for ScF$_3$ }
\end{figure}
The results for the symmetry-adapted analysis are shown in Fig \ref{f2} (middle and bottom). The distortions can be classed into 3 general types: rigid unit modes, consisting of coherent rotations of the octahedra; semi-rigid "scissoring" modes, where there is a scissoring of some of the M-F bonds within the octahedra, and bond-stretching modes, where some M-F bond lengths change. Most irreps in this analysis only have one distortion associated with them, although there are a few with more. There is a good degree of consistency between the two compounds; both have two "bands" of modes, one that fits well and one that fits poorly. The band with a greater weighted R-factor in both compounds consist of the same irreps (R\textsubscript{3}\textsuperscript{-}, R\textsubscript{5}\textsuperscript{+}, X\textsubscript{3}\textsuperscript{-}, M\textsubscript{2}\textsuperscript{-}, M\textsubscript{3}\textsuperscript{+}, M\textsubscript{4}\textsuperscript{+} (and R\textsubscript{2}\textsuperscript{-} in ScF\textsubscript{3})), all of which have distortions with a bond-stretching character. The rest of the irreps, in the band that fits the data well, have at least one distortion associated with them that has a rigid unit (M$_2^{+}$ and R$_5^{-}$) or scissoring mode character. There are 4 zone boundary irreps that consistently have the lowest weighted R-factor for both compounds for the majority of temperatures - X$_{5}^{+}$, X$_{5}^{-}$, M$_{5}^{+}$ and M$_{5}^{-}$. All of these irreps have one distortion associated with them that is of scissoring mode character. A depiction of the effect of these modes on the structure of CaZrF$_6$ is shown in Fig \ref{f7}. The $\mathrm{\Gamma}_5^{-}$ irrep also fits well, especially in the refinements that go out to $\SI{30}{\angstrom}$. The displacements associated with this irrep are also of a scissoring mode character, however despite the low R$_{wp}$, the mode amplitudes are consistently small, hence most of the analysis is focused on X$_{5}^{+}$, X$_{5}^{-}$, M$_{5}^{+}$ and M$_{5}^{-}$. The weighted mean amplitude over all refinements at each temperature for these irreps have been calculated and are shown in Fig. \ref{f2} (bottom), the weighting being given by a Boltzmann distribution $ w =  exp[(R_{global} - R_{wp})/\sigma] $, where $R_{global}$ is the minimum weighted R-factor achieved across all refinements and all temperatures for the relevant compound and $\sigma$  is  the value of a meaningful difference in the weighted R-factor, taken to be $0.1 \% $. $R_{global}$ is taken to be $9 \% $ for both compounds. The amplitude of these modes (X$_{5}^{+}$, X$_{5}^{-}$, M$_{5}^{+}$ and M$_{5}^{-}$) are consistently higher for CaZrF$_6$ than for ScF$_3$ - this coincides well with the more significant distortion away from the average structure for CaZrF$_6$, as seen in the mean M-F-M bond angles, and the greater magnitude of NTE reported in the literature. These modes also fit significantly better than the RUMs (M$_{2}^{+}$ and R$_{5}^{-}$). These best-fitting irreps (X$_{5}^{+}$, X$_{5}^{-}$, M$_{5}^{+}$ and M$_{5}^{-}$) are all two-dimensional and all have 3 k-vectors, therefore the OPDs have been constrained as described in section II.B to allow for a fairer comparison with the RUMs, which have fewer degrees of freedom associated with them. For the ScF$_3$, the unconstrained R$_{5}^{-}$ (which is associated with the out-of-phase octahdral tilts observed in other metal trifluorides) has a similar quality of fit to the constrained X$_{5}^{+}$, X$_{5}^{-}$ and M$_{5}^{+}$ at lower temperatures and consistently performs better than M$_{5}^{-}$ (Fig \ref{f3}). This suggests that a combination of both the rigid unit and scissoring modes are responsible for NTE, which agrees with previous molecular dynamics studies of these materials\cite{Schick2016}. In this study the authors argue that correlated dynamics of flexible polyhedra result in a greater degree of NTE than purely rigid unit dynamics. However for CaZrF$_6$, we find the constrained scissoring modes, with the exception of M$_{5}^{-}$, consistently perform better than the RUMs. The RUMs also start to perform increasingly poorly as the temperature is raised above 100 K, suggesting that the thermal expansion in CaZrF$_6$ at higher temperatures may well be dominated by contributions from these scissoring modes. The increasing R$_{wp}$ of the RUMs as temperature is increased, and the contrasting decrease in R$_{wp}$ seen for the scissoring modes, tallies well with the phonon dispersion curves of both compounds\cite{Gupta2018,Keith2011}. These show that the scissoring modes are slightly higher in energy than the RUMs, so the scissoring modes will become more active at higher temperatures.

As discussed earlier, the different charges on the two cations in CaZrF$_6$ result in a need to refine the octahedral breathing mode, transforming as the R$_2^{-}$ irrep, alongside the other distortion modes in order to facilitate a more direct comparison to ScF$_3$. In the average structure of CaZrF$_6$, this breathing mode is frozen in, lowering the symmetry from $Pm\bar{3}m$ to $Fm\bar{3}m$. This also has the effect of mixing the characters of some of the irreps such that the associated atomic displacements now transform as the same irrep. For example, the X$_{5}^{+}$ and M$_{5}^{-}$ irreps of $Pm\bar{3}m$ correspond to the X$_{5}^{-}$ irrep of $Fm\bar{3}m$ and X$_{5}^{-}$ and M$_{5}^{+}$ correspond to X$_{5}^{+}$. To determine whether this mixing of characters has any effect on the observed local structure of CaZrF$_6$, the constrained OPD X$_{5}^{+}$ and M$_{5}^{-}$ modes were refined together (hereafter referred to as X$_{5}^{+}$$\oplus$M$_{5}^{-}$). This gave a significant improvement to the quality of the fit (Figs \ref{f3} and \ref{f4}). To determine whether this coupling is a significant effect, results are compared to a two-phase model, in which modes transforming as different irreps are refined in separate phases. Hereafter these two models will be referred to as the "coupled" model (denoted with $\oplus$) and the "2-phase" model (denoted with \&). The coupled modes have a significantly better R-factor above 100 K, but fit worse than the 2-phase refinement below this temperature.  The same comparisons are also done for ScF$_3$, where any coupling between phonons of these characters should only arise from anharmonic interactions. In contrast to CaZrF$_6$, which shows a clear preference for coupling between X$_{5}^{+}$ and M$_{5}^{-}$, no evidence of such coupling and hence an anharmonic interaction is seen here for ScF$_3$. This suggests that whilst these scissoring modes are important in determining the local structure of ScF$_3$, any anharmonic coupling between them has little influence on the lattice dynamics that drive NTE. A similar comparison is made for X$_{5}^{-}$/M$_{5}^{+}$, however the 2-phase refinements consistently fit better than the coupled model for both compounds. This may be due to both distortions locally having the same character (Eu of point group $m\bar{3}m$) with respect to the MF$_6$ octahedra, making coupling unfavourable.

Next we investigate if the similar quality of fits of the scissoring modes X$_{5}^{+}$, X$_{5}^{-}$ and M$_{5}^{+}$ and the rigid unit mode R$_{5}^{-}$ could be indicative that the two types of distortion are  cooperatively coupled to produce the observed NTE. To test this hypothesis, we explore 2 scenarios; whether this observation is simply due to the dynamic distortions occurring in different sample volumes or at different times to each other or a coupled model which implies that significant anharmonic coupling between these modes is occurring. For both materials, the X$_{5}^{+}$/R$_{5}^{-}$ refinements show a similar sort of behaviour to the X$_{5}^{+}$$\oplus$M$_{5}^{-}$ refinements in CaZrF$_6$, in that the refinements of the coupled modes perform worse than the 2-phase refinements at lower temperatures, but soon crossover to show an improved fit, although the results for CaZrF$_6$ are not robust. Since by the symmetry lowering of the rock-salt ordering in CaZrF$_6$, X$_{5}^{+}$ and M$_{5}^{-}$ are allowed to couple, and we have shown our analysis to be sensitive to this coupling, the results in Fig. \ref{f6} are indicative that there is coupling between the X$_{5}^{+}$ and R$_{5}^{-}$ modes. However, as by symmetry coupling in X$_{5}^{+}$$\oplus$R$_{5}^{-}$ is not permitted on its own, we construct a coupled distortion that forms an invariant in the free energy expansion by inclusion of the M$_{5}^{-}$ irrep. The X$_{5}^{+}$ and M$_{5}^{-}$ OPDs in this refinement are still restricted, resulting in 3 more parameters than the X$_{5}^{+}$$\oplus$M$_{5}^{-}$ refinements, but much improved fits (Figs \ref{f3} and \ref{f6}). This model results in a very good agreement with the data (Fig. \ref{f6}).  

A very recent analysis of ScF$_3$ neutron PDF data using the reverse Monte Carlo (RMC) method\cite{Dove2019} similarly concludes that it is a combination between structural flexibility and RUMs that causes the NTE in the compound. Dove \textit{et al.} argue that the flexibility of the structure allows RUMs and RUM-like modes to occupy a larger volume in reciprocal space, meaning they give a greater contribution to the overall thermal expansion behaviour, compared to entirely rigid structures. Our results here echo this conclusion and underline the dominant contribution of scissoring modes in describing the fluctuations from the average symmetry. Additionally in the work of Dove \textit{et al.}, geometric algebra is used to quantify the proportion of the motion of the atoms in ScF$_3$ originating from correlated whole-body octahedral motion, deformations of the F-Sc-F right angles and changes to the Sc-F bond length. This analysis resulted in a ratio of approximately 7:2:1 of bends:rotations:stretches. The X$_{5}^{+}$\&R$_{5}^{-}$ and X$_{5}^{-}$\&R$_{5}^{-}$ 2-phase refinements desribed previously give a similar ratio of bends:rotations, approximately 8:2, although the contribution from stretches is negligible ($< 1 \% $ of the total motion). The X$_{5}^{+}$\&R$_{5}^{-}$ refinements for CaZrF$_6$ give an approximately 7:3 ratio of bends:rotations, again with a negligible contribution from stretches. There is hence a high degree of consistency between results derived via big box RMC methods and those of our symmetry motivated approach here. A different analysis of neutron PDF data of ScF$_3$, performed by Wendt \textit{et al.}\cite{Wendt2019}, models the F atoms as being randomly positioned on a torus-shaped gaussian distribution around the F sites in the average structure, with no correlation between neighbouring F atoms, in a similar fashion to entropic elasticity in polymers. The model reproduces the observed NTE behaviour and F-F distribution up to $\approx$ 700 K. It shows how important flexibility of Sc-F-Sc linkages are in this material, a fact consistent with our findings here, however it fails to account for the full range of NTE in the material. The previously discussed RMC model shows that at least a small fraction of the motion of F atoms in the material can be accounted for with correlated rigid-unit type distortions, results which are compatible with our symmetry based analysis of the X-ray PDF data.

In summary, we have shown via a symmetry motivated real-space analysis of PDF data that the most significant distortions in these ReO$_3$-like NTE materials are scissoring modes, which involve scissoring of the MF$_6$ octahedral bond angles. These modes have a greater amplitude in CaZrF$_6$ than ScF$_3$, which corresponds well to the greater magnitude of NTE reported in the literature for the former. Coupling between these modes and the rigid unit modes has been shown to be active and the likely origin of unusually high NTE in these structures. 

\section*{Acknowledgements}
T. A. Bird thanks EPSRC for a PhD studentship through the EPSRC Centre for Doctoral Training in Molecular Analytical Science, grant number EP/L015307/1.\\

M. S. Senn acknowledges the Royal Commission for the Exhibition of 1851 and the Royal Society for fellowships.\\

We acknowledge DESY (Hamburg, Germany), a member of the Helmholtz Association HGF, for the provision of experimental facilities. Parts of this research were carried out at PETRA III.\\

This research used resources of the Advanced Photon Source, a U.S. Department of Energy (DOE) Office of Science User Facility operated for the DOE Office of Science by Argonne National Laboratory under Contract No. DE-AC02-06CH11357. We acknowledge the measurement of PDF data by Dr. Y. Ren.\\

Samples were characterised using the I11 beamline at Diamond Light Source before total scattering experiments were performed. Access to this beamline was granted via Block Allocation Group EE18786.

\subsection*{Notes}

The ScF$_3$ data for this study is available as a supporting data set at \href{https://doi.org/10.6084/m9.figshare.11605278.v1}{https://doi.org/10.6084/m9.figshare.11605278.v1}

\appendix
\bibliographystyle{apsrev4-1}
\bibliography{nte_refs3.bib}

\begin{thebibliography}{38}%
\makeatletter
\providecommand \@ifxundefined [1]{%
 \@ifx{#1\undefined}
}%
\providecommand \@ifnum [1]{%
 \ifnum #1\expandafter \@firstoftwo
 \else \expandafter \@secondoftwo
 \fi
}%
\providecommand \@ifx [1]{%
 \ifx #1\expandafter \@firstoftwo
 \else \expandafter \@secondoftwo
 \fi
}%
\providecommand \natexlab [1]{#1}%
\providecommand \enquote  [1]{``#1''}%
\providecommand \bibnamefont  [1]{#1}%
\providecommand \bibfnamefont [1]{#1}%
\providecommand \citenamefont [1]{#1}%
\providecommand \href@noop [0]{\@secondoftwo}%
\providecommand \href [0]{\begingroup \@sanitize@url \@href}%
\providecommand \@href[1]{\@@startlink{#1}\@@href}%
\providecommand \@@href[1]{\endgroup#1\@@endlink}%
\providecommand \@sanitize@url [0]{\catcode `\\12\catcode `\$12\catcode
  `\&12\catcode `\#12\catcode `\^12\catcode `\_12\catcode `\%12\relax}%
\providecommand \@@startlink[1]{}%
\providecommand \@@endlink[0]{}%
\providecommand \url  [0]{\begingroup\@sanitize@url \@url }%
\providecommand \@url [1]{\endgroup\@href {#1}{\urlprefix }}%
\providecommand \urlprefix  [0]{URL }%
\providecommand \Eprint [0]{\href }%
\providecommand \doibase [0]{http://dx.doi.org/}%
\providecommand \selectlanguage [0]{\@gobble}%
\providecommand \bibinfo  [0]{\@secondoftwo}%
\providecommand \bibfield  [0]{\@secondoftwo}%
\providecommand \translation [1]{[#1]}%
\providecommand \BibitemOpen [0]{}%
\providecommand \bibitemStop [0]{}%
\providecommand \bibitemNoStop [0]{.\EOS\space}%
\providecommand \EOS [0]{\spacefactor3000\relax}%
\providecommand \BibitemShut  [1]{\csname bibitem#1\endcsname}%
\let\auto@bib@innerbib\@empty
\bibitem [{\citenamefont {Evans}\ \emph {et~al.}(1996)\citenamefont {Evans},
  \citenamefont {Mary}, \citenamefont {Vogt}, \citenamefont {Subramanian},\
  and\ \citenamefont {Sleight}}]{Evans1996}%
  \BibitemOpen
  \bibfield  {author} {\bibinfo {author} {\bibfnamefont {J.~S.}\ \bibnamefont
  {Evans}}, \bibinfo {author} {\bibfnamefont {T.~A.}\ \bibnamefont {Mary}},
  \bibinfo {author} {\bibfnamefont {T.}~\bibnamefont {Vogt}}, \bibinfo {author}
  {\bibfnamefont {M.~A.}\ \bibnamefont {Subramanian}}, \ and\ \bibinfo {author}
  {\bibfnamefont {A.~W.}\ \bibnamefont {Sleight}},\ }\href {\doibase
  10.1021/cm9602959} {\bibfield  {journal} {\bibinfo  {journal} {Chemistry of
  Materials}\ }\textbf {\bibinfo {volume} {8}},\ \bibinfo {pages} {2809}
  (\bibinfo {year} {1996})}\BibitemShut {NoStop}%
\bibitem [{\citenamefont {Dapiaggi}\ \emph {et~al.}(2003)\citenamefont
  {Dapiaggi}, \citenamefont {Tiano}, \citenamefont {Artioli}, \citenamefont
  {Sanson},\ and\ \citenamefont {Fornasini}}]{Dapiaggi2003}%
  \BibitemOpen
  \bibfield  {author} {\bibinfo {author} {\bibfnamefont {M.}~\bibnamefont
  {Dapiaggi}}, \bibinfo {author} {\bibfnamefont {W.}~\bibnamefont {Tiano}},
  \bibinfo {author} {\bibfnamefont {G.}~\bibnamefont {Artioli}}, \bibinfo
  {author} {\bibfnamefont {A.}~\bibnamefont {Sanson}}, \ and\ \bibinfo {author}
  {\bibfnamefont {P.}~\bibnamefont {Fornasini}},\ }\href {\doibase
  10.1016/S0168-583X(02)01682-8} {\bibfield  {journal} {\bibinfo  {journal}
  {Nuclear Instruments and Methods in Physics Research, Section B: Beam
  Interactions with Materials and Atoms}\ }\textbf {\bibinfo {volume} {200}},\
  \bibinfo {pages} {231} (\bibinfo {year} {2003})}\BibitemShut {NoStop}%
\bibitem [{\citenamefont {Chatterji}\ \emph {et~al.}(2008)\citenamefont
  {Chatterji}, \citenamefont {Henry}, \citenamefont {Mittal},\ and\
  \citenamefont {Chaplot}}]{Chatterji2008}%
  \BibitemOpen
  \bibfield  {author} {\bibinfo {author} {\bibfnamefont {T.}~\bibnamefont
  {Chatterji}}, \bibinfo {author} {\bibfnamefont {P.~F.}\ \bibnamefont
  {Henry}}, \bibinfo {author} {\bibfnamefont {R.}~\bibnamefont {Mittal}}, \
  and\ \bibinfo {author} {\bibfnamefont {S.~L.}\ \bibnamefont {Chaplot}},\
  }\href {\doibase 10.1103/PhysRevB.78.134105} {\bibfield  {journal} {\bibinfo
  {journal} {Physical Review B - Condensed Matter and Materials Physics}\
  }\textbf {\bibinfo {volume} {78}},\ \bibinfo {pages} {3} (\bibinfo {year}
  {2008})}\BibitemShut {NoStop}%
\bibitem [{\citenamefont {Chatterji}\ \emph {et~al.}(2009)\citenamefont
  {Chatterji}, \citenamefont {Hansen}, \citenamefont {Brunelli},\ and\
  \citenamefont {Henry}}]{Chatterji2009}%
  \BibitemOpen
  \bibfield  {author} {\bibinfo {author} {\bibfnamefont {T.}~\bibnamefont
  {Chatterji}}, \bibinfo {author} {\bibfnamefont {T.~C.}\ \bibnamefont
  {Hansen}}, \bibinfo {author} {\bibfnamefont {M.}~\bibnamefont {Brunelli}}, \
  and\ \bibinfo {author} {\bibfnamefont {P.~F.}\ \bibnamefont {Henry}},\ }\href
  {\doibase 10.1063/1.3155191} {\bibfield  {journal} {\bibinfo  {journal}
  {Applied Physics Letters}\ }\textbf {\bibinfo {volume} {94}},\ \bibinfo
  {pages} {3} (\bibinfo {year} {2009})}\BibitemShut {NoStop}%
\bibitem [{\citenamefont {Grobler}\ \emph {et~al.}(2013)\citenamefont
  {Grobler}, \citenamefont {Smith}, \citenamefont {Bhatt}, \citenamefont
  {Herbert},\ and\ \citenamefont {Barbour}}]{Grobler2013}%
  \BibitemOpen
  \bibfield  {author} {\bibinfo {author} {\bibfnamefont {I.}~\bibnamefont
  {Grobler}}, \bibinfo {author} {\bibfnamefont {V.~J.}\ \bibnamefont {Smith}},
  \bibinfo {author} {\bibfnamefont {P.~M.}\ \bibnamefont {Bhatt}}, \bibinfo
  {author} {\bibfnamefont {S.~A.}\ \bibnamefont {Herbert}}, \ and\ \bibinfo
  {author} {\bibfnamefont {L.~J.}\ \bibnamefont {Barbour}},\ }\href {\doibase
  10.1021/ja401671p} {\bibfield  {journal} {\bibinfo  {journal} {Journal of the
  American Chemical Society}\ }\textbf {\bibinfo {volume} {135}},\ \bibinfo
  {pages} {6411} (\bibinfo {year} {2013})}\BibitemShut {NoStop}%
\bibitem [{\citenamefont {Lock}\ \emph {et~al.}(2010)\citenamefont {Lock},
  \citenamefont {Wu}, \citenamefont {Christensen}, \citenamefont {Cameron},
  \citenamefont {Peterson}, \citenamefont {Bridgeman}, \citenamefont {Kepert},\
  and\ \citenamefont {Iversen}}]{Lock2010}%
  \BibitemOpen
  \bibfield  {author} {\bibinfo {author} {\bibfnamefont {N.}~\bibnamefont
  {Lock}}, \bibinfo {author} {\bibfnamefont {Y.}~\bibnamefont {Wu}}, \bibinfo
  {author} {\bibfnamefont {M.}~\bibnamefont {Christensen}}, \bibinfo {author}
  {\bibfnamefont {L.~J.}\ \bibnamefont {Cameron}}, \bibinfo {author}
  {\bibfnamefont {V.~K.}\ \bibnamefont {Peterson}}, \bibinfo {author}
  {\bibfnamefont {A.~J.}\ \bibnamefont {Bridgeman}}, \bibinfo {author}
  {\bibfnamefont {C.~J.}\ \bibnamefont {Kepert}}, \ and\ \bibinfo {author}
  {\bibfnamefont {B.~B.}\ \bibnamefont {Iversen}},\ }\href {\doibase
  10.1021/jp103212z} {\bibfield  {journal} {\bibinfo  {journal} {Journal of
  Physical Chemistry C}\ }\textbf {\bibinfo {volume} {114}},\ \bibinfo {pages}
  {16181} (\bibinfo {year} {2010})}\BibitemShut {NoStop}%
\bibitem [{\citenamefont {Dove}\ and\ \citenamefont {Fang}(2016)}]{Dove2016}%
  \BibitemOpen
  \bibfield  {author} {\bibinfo {author} {\bibfnamefont {M.~T.}\ \bibnamefont
  {Dove}}\ and\ \bibinfo {author} {\bibfnamefont {H.}~\bibnamefont {Fang}},\
  }\href {\doibase 10.1088/0034-4885/79/6/066503} {\enquote {\bibinfo {title}
  {{Negative thermal expansion and associated anomalous physical properties:
  Review of the lattice dynamics theoretical foundation}},}\ } (\bibinfo {year}
  {2016})\BibitemShut {NoStop}%
\bibitem [{\citenamefont {Barrera}\ \emph {et~al.}(2005)\citenamefont
  {Barrera}, \citenamefont {Bruno}, \citenamefont {Barron},\ and\ \citenamefont
  {Allan}}]{Barrera2005}%
  \BibitemOpen
  \bibfield  {author} {\bibinfo {author} {\bibfnamefont {G.~D.}\ \bibnamefont
  {Barrera}}, \bibinfo {author} {\bibfnamefont {J.~A.~O.}\ \bibnamefont
  {Bruno}}, \bibinfo {author} {\bibfnamefont {T.~H.~K.}\ \bibnamefont
  {Barron}}, \ and\ \bibinfo {author} {\bibfnamefont {N.~L.}\ \bibnamefont
  {Allan}},\ }\href {\doibase 10.1088/0953-8984/17/4/R03} {\bibfield  {journal}
  {\bibinfo  {journal} {J. Phys. Condens. Matter}\ }\textbf {\bibinfo {volume}
  {17}},\ \bibinfo {pages} {217} (\bibinfo {year} {2005})}\BibitemShut
  {NoStop}%
\bibitem [{\citenamefont {Greve}\ \emph {et~al.}(2010)\citenamefont {Greve},
  \citenamefont {Martin}, \citenamefont {Lee}, \citenamefont {Chupas},
  \citenamefont {Chapman},\ and\ \citenamefont {Wilkinson}}]{Greve2010}%
  \BibitemOpen
  \bibfield  {author} {\bibinfo {author} {\bibfnamefont {B.~K.}\ \bibnamefont
  {Greve}}, \bibinfo {author} {\bibfnamefont {K.~L.}\ \bibnamefont {Martin}},
  \bibinfo {author} {\bibfnamefont {P.~L.}\ \bibnamefont {Lee}}, \bibinfo
  {author} {\bibfnamefont {P.~J.}\ \bibnamefont {Chupas}}, \bibinfo {author}
  {\bibfnamefont {K.~W.}\ \bibnamefont {Chapman}}, \ and\ \bibinfo {author}
  {\bibfnamefont {A.~P.}\ \bibnamefont {Wilkinson}},\ }\href {\doibase
  10.1021/ja106711v} {\bibfield  {journal} {\bibinfo  {journal} {Journal of the
  American Chemical Society}\ }\textbf {\bibinfo {volume} {132}},\ \bibinfo
  {pages} {15496} (\bibinfo {year} {2010})}\BibitemShut {NoStop}%
\bibitem [{\citenamefont {Handunkanda}\ \emph {et~al.}(2015)\citenamefont
  {Handunkanda}, \citenamefont {Curry}, \citenamefont {Voronov}, \citenamefont
  {Said}, \citenamefont {Guzm{\'{a}}n-Verri}, \citenamefont {Brierley},
  \citenamefont {Littlewood},\ and\ \citenamefont {Hancock}}]{Handunkanda2015}%
  \BibitemOpen
  \bibfield  {author} {\bibinfo {author} {\bibfnamefont {S.~U.}\ \bibnamefont
  {Handunkanda}}, \bibinfo {author} {\bibfnamefont {E.~B.}\ \bibnamefont
  {Curry}}, \bibinfo {author} {\bibfnamefont {V.}~\bibnamefont {Voronov}},
  \bibinfo {author} {\bibfnamefont {A.~H.}\ \bibnamefont {Said}}, \bibinfo
  {author} {\bibfnamefont {G.~G.}\ \bibnamefont {Guzm{\'{a}}n-Verri}}, \bibinfo
  {author} {\bibfnamefont {R.~T.}\ \bibnamefont {Brierley}}, \bibinfo {author}
  {\bibfnamefont {P.~B.}\ \bibnamefont {Littlewood}}, \ and\ \bibinfo {author}
  {\bibfnamefont {J.~N.}\ \bibnamefont {Hancock}},\ }\href {\doibase
  10.1103/PhysRevB.92.134101} {\bibfield  {journal} {\bibinfo  {journal}
  {Physical Review B - Condensed Matter and Materials Physics}\ }\textbf
  {\bibinfo {volume} {92}} (\bibinfo {year} {2015}),\
  10.1103/PhysRevB.92.134101}\BibitemShut {NoStop}%
\bibitem [{\citenamefont {Hancock}\ \emph {et~al.}(2015)\citenamefont
  {Hancock}, \citenamefont {Chapman}, \citenamefont {Halder}, \citenamefont
  {Morelock}, \citenamefont {Kaplan}, \citenamefont {Gallington}, \citenamefont
  {Bongiorno}, \citenamefont {Han}, \citenamefont {Zhou},\ and\ \citenamefont
  {Wilkinson}}]{Hancock2015}%
  \BibitemOpen
  \bibfield  {author} {\bibinfo {author} {\bibfnamefont {J.~C.}\ \bibnamefont
  {Hancock}}, \bibinfo {author} {\bibfnamefont {K.~W.}\ \bibnamefont
  {Chapman}}, \bibinfo {author} {\bibfnamefont {G.~J.}\ \bibnamefont {Halder}},
  \bibinfo {author} {\bibfnamefont {C.~R.}\ \bibnamefont {Morelock}}, \bibinfo
  {author} {\bibfnamefont {B.~S.}\ \bibnamefont {Kaplan}}, \bibinfo {author}
  {\bibfnamefont {L.~C.}\ \bibnamefont {Gallington}}, \bibinfo {author}
  {\bibfnamefont {A.}~\bibnamefont {Bongiorno}}, \bibinfo {author}
  {\bibfnamefont {C.}~\bibnamefont {Han}}, \bibinfo {author} {\bibfnamefont
  {S.}~\bibnamefont {Zhou}}, \ and\ \bibinfo {author} {\bibfnamefont {A.~P.}\
  \bibnamefont {Wilkinson}},\ }\href {\doibase 10.1021/acs.chemmater.5b00662}
  {\bibfield  {journal} {\bibinfo  {journal} {Chemistry of Materials}\ }\textbf
  {\bibinfo {volume} {27}},\ \bibinfo {pages} {3912} (\bibinfo {year}
  {2015})}\BibitemShut {NoStop}%
\bibitem [{\citenamefont {Hu}\ \emph {et~al.}(2016{\natexlab{a}})\citenamefont
  {Hu}, \citenamefont {Chen}, \citenamefont {Sanson}, \citenamefont {Wu},
  \citenamefont {{Guglieri Rodriguez}}, \citenamefont {Olivi}, \citenamefont
  {Ren}, \citenamefont {Fan}, \citenamefont {Deng},\ and\ \citenamefont
  {Xing}}]{Hu2016}%
  \BibitemOpen
  \bibfield  {author} {\bibinfo {author} {\bibfnamefont {L.}~\bibnamefont
  {Hu}}, \bibinfo {author} {\bibfnamefont {J.}~\bibnamefont {Chen}}, \bibinfo
  {author} {\bibfnamefont {A.}~\bibnamefont {Sanson}}, \bibinfo {author}
  {\bibfnamefont {H.}~\bibnamefont {Wu}}, \bibinfo {author} {\bibfnamefont
  {C.}~\bibnamefont {{Guglieri Rodriguez}}}, \bibinfo {author} {\bibfnamefont
  {L.}~\bibnamefont {Olivi}}, \bibinfo {author} {\bibfnamefont
  {Y.}~\bibnamefont {Ren}}, \bibinfo {author} {\bibfnamefont {L.}~\bibnamefont
  {Fan}}, \bibinfo {author} {\bibfnamefont {J.}~\bibnamefont {Deng}}, \ and\
  \bibinfo {author} {\bibfnamefont {X.}~\bibnamefont {Xing}},\ }\href {\doibase
  10.1021/jacs.6b02370} {\bibfield  {journal} {\bibinfo  {journal} {Journal of
  the American Chemical Society}\ }\textbf {\bibinfo {volume} {138}},\ \bibinfo
  {pages} {8320} (\bibinfo {year} {2016}{\natexlab{a}})}\BibitemShut {NoStop}%
\bibitem [{\citenamefont {Gupta}\ \emph {et~al.}(2018)\citenamefont {Gupta},
  \citenamefont {Singh}, \citenamefont {Mittal},\ and\ \citenamefont
  {Chaplot}}]{Gupta2018}%
  \BibitemOpen
  \bibfield  {author} {\bibinfo {author} {\bibfnamefont {M.}~\bibnamefont
  {Gupta}}, \bibinfo {author} {\bibfnamefont {B.}~\bibnamefont {Singh}},
  \bibinfo {author} {\bibfnamefont {R.}~\bibnamefont {Mittal}}, \ and\ \bibinfo
  {author} {\bibfnamefont {S.~L.}\ \bibnamefont {Chaplot}},\ }\href@noop {}
  {\bibfield  {journal} {\bibinfo  {journal} {Physical Review B - Condensed
  Matter and Materials Physics}\ }\textbf {\bibinfo {volume} {98}},\ \bibinfo
  {pages} {14301} (\bibinfo {year} {2018})}\BibitemShut {NoStop}%
\bibitem [{\citenamefont {Tucker}\ \emph {et~al.}(2005)\citenamefont {Tucker},
  \citenamefont {Goodwin}, \citenamefont {Dove}, \citenamefont {Keen},
  \citenamefont {Wells},\ and\ \citenamefont {Evans}}]{Tucker2005}%
  \BibitemOpen
  \bibfield  {author} {\bibinfo {author} {\bibfnamefont {M.~G.}\ \bibnamefont
  {Tucker}}, \bibinfo {author} {\bibfnamefont {A.~L.}\ \bibnamefont {Goodwin}},
  \bibinfo {author} {\bibfnamefont {M.~T.}\ \bibnamefont {Dove}}, \bibinfo
  {author} {\bibfnamefont {D.~A.}\ \bibnamefont {Keen}}, \bibinfo {author}
  {\bibfnamefont {S.~A.}\ \bibnamefont {Wells}}, \ and\ \bibinfo {author}
  {\bibfnamefont {J.~S.~O.}\ \bibnamefont {Evans}},\ }\href {\doibase
  10.1103/PhysRevLett.95.255501} {\bibfield  {journal} {\bibinfo  {journal}
  {Physical Review Letters}\ }\textbf {\bibinfo {volume} {95}},\ \bibinfo
  {pages} {8} (\bibinfo {year} {2005})}\BibitemShut {NoStop}%
\bibitem [{\citenamefont {Sanson}\ \emph {et~al.}(2016)\citenamefont {Sanson},
  \citenamefont {Giarola}, \citenamefont {Mariotto}, \citenamefont {Hu},
  \citenamefont {Chen},\ and\ \citenamefont {Xing}}]{Sanson2016}%
  \BibitemOpen
  \bibfield  {author} {\bibinfo {author} {\bibfnamefont {A.}~\bibnamefont
  {Sanson}}, \bibinfo {author} {\bibfnamefont {M.}~\bibnamefont {Giarola}},
  \bibinfo {author} {\bibfnamefont {G.}~\bibnamefont {Mariotto}}, \bibinfo
  {author} {\bibfnamefont {L.}~\bibnamefont {Hu}}, \bibinfo {author}
  {\bibfnamefont {J.}~\bibnamefont {Chen}}, \ and\ \bibinfo {author}
  {\bibfnamefont {X.}~\bibnamefont {Xing}},\ }\href {\doibase
  10.1016/j.matchemphys.2016.05.067} {\bibfield  {journal} {\bibinfo  {journal}
  {Materials Chemistry and Physics}\ }\textbf {\bibinfo {volume} {180}},\
  \bibinfo {pages} {213} (\bibinfo {year} {2016})}\BibitemShut {NoStop}%
\bibitem [{\citenamefont {Sennova}\ \emph {et~al.}(2007)\citenamefont
  {Sennova}, \citenamefont {Bubnova}, \citenamefont {Shepelev}, \citenamefont
  {Filatov},\ and\ \citenamefont {Yakovleva}}]{Sennova2007}%
  \BibitemOpen
  \bibfield  {author} {\bibinfo {author} {\bibfnamefont {N.}~\bibnamefont
  {Sennova}}, \bibinfo {author} {\bibfnamefont {R.}~\bibnamefont {Bubnova}},
  \bibinfo {author} {\bibfnamefont {J.}~\bibnamefont {Shepelev}}, \bibinfo
  {author} {\bibfnamefont {S.}~\bibnamefont {Filatov}}, \ and\ \bibinfo
  {author} {\bibfnamefont {O.}~\bibnamefont {Yakovleva}},\ }\href {\doibase
  10.1016/j.jallcom.2006.03.049} {\bibfield  {journal} {\bibinfo  {journal}
  {Journal of Alloys and Compounds}\ }\textbf {\bibinfo {volume} {428}},\
  \bibinfo {pages} {290} (\bibinfo {year} {2007})}\BibitemShut {NoStop}%
\bibitem [{\citenamefont {Senyshyn}\ \emph {et~al.}(2010)\citenamefont
  {Senyshyn}, \citenamefont {Schwarz}, \citenamefont {Lorenz}, \citenamefont
  {Adamiv}, \citenamefont {Burak}, \citenamefont {Banys}, \citenamefont
  {Grigalaitis}, \citenamefont {Vasylechko}, \citenamefont {Ehrenberg},\ and\
  \citenamefont {Fuess}}]{Senyshyn2010}%
  \BibitemOpen
  \bibfield  {author} {\bibinfo {author} {\bibfnamefont {A.}~\bibnamefont
  {Senyshyn}}, \bibinfo {author} {\bibfnamefont {B.}~\bibnamefont {Schwarz}},
  \bibinfo {author} {\bibfnamefont {T.}~\bibnamefont {Lorenz}}, \bibinfo
  {author} {\bibfnamefont {V.~T.}\ \bibnamefont {Adamiv}}, \bibinfo {author}
  {\bibfnamefont {Y.~V.}\ \bibnamefont {Burak}}, \bibinfo {author}
  {\bibfnamefont {J.}~\bibnamefont {Banys}}, \bibinfo {author} {\bibfnamefont
  {R.}~\bibnamefont {Grigalaitis}}, \bibinfo {author} {\bibfnamefont
  {L.}~\bibnamefont {Vasylechko}}, \bibinfo {author} {\bibfnamefont
  {H.}~\bibnamefont {Ehrenberg}}, \ and\ \bibinfo {author} {\bibfnamefont
  {H.}~\bibnamefont {Fuess}},\ }\href {\doibase 10.1063/1.3504244} {\bibfield
  {journal} {\bibinfo  {journal} {Journal of Applied Physics}\ }\textbf
  {\bibinfo {volume} {108}} (\bibinfo {year} {2010}),\
  10.1063/1.3504244}\BibitemShut {NoStop}%
\bibitem [{\citenamefont {Hu}\ \emph {et~al.}(2016{\natexlab{b}})\citenamefont
  {Hu}, \citenamefont {Chen}, \citenamefont {Xu}, \citenamefont {Wang},
  \citenamefont {Han}, \citenamefont {Ren}, \citenamefont {Pan}, \citenamefont
  {Rong}, \citenamefont {Huang}, \citenamefont {Deng}, \citenamefont {Li},\
  and\ \citenamefont {Xing}}]{Hu2016a}%
  \BibitemOpen
  \bibfield  {author} {\bibinfo {author} {\bibfnamefont {L.}~\bibnamefont
  {Hu}}, \bibinfo {author} {\bibfnamefont {J.}~\bibnamefont {Chen}}, \bibinfo
  {author} {\bibfnamefont {J.}~\bibnamefont {Xu}}, \bibinfo {author}
  {\bibfnamefont {N.}~\bibnamefont {Wang}}, \bibinfo {author} {\bibfnamefont
  {F.}~\bibnamefont {Han}}, \bibinfo {author} {\bibfnamefont {Y.}~\bibnamefont
  {Ren}}, \bibinfo {author} {\bibfnamefont {Z.}~\bibnamefont {Pan}}, \bibinfo
  {author} {\bibfnamefont {Y.}~\bibnamefont {Rong}}, \bibinfo {author}
  {\bibfnamefont {R.}~\bibnamefont {Huang}}, \bibinfo {author} {\bibfnamefont
  {J.}~\bibnamefont {Deng}}, \bibinfo {author} {\bibfnamefont {L.}~\bibnamefont
  {Li}}, \ and\ \bibinfo {author} {\bibfnamefont {X.}~\bibnamefont {Xing}},\
  }\href {\doibase 10.1021/jacs.6b08746} {\bibfield  {journal} {\bibinfo
  {journal} {Journal of the American Chemical Society}\ }\textbf {\bibinfo
  {volume} {138}},\ \bibinfo {pages} {14530} (\bibinfo {year}
  {2016}{\natexlab{b}})}\BibitemShut {NoStop}%
\bibitem [{\citenamefont {Keith}\ \emph {et~al.}(2011)\citenamefont {Keith},
  \citenamefont {Tracy}, \citenamefont {Abernathy}, \citenamefont {Fultz},
  \citenamefont {Li}, \citenamefont {Tang},\ and\ \citenamefont
  {Mun}}]{Keith2011}%
  \BibitemOpen
  \bibfield  {author} {\bibinfo {author} {\bibfnamefont {J.~B.}\ \bibnamefont
  {Keith}}, \bibinfo {author} {\bibfnamefont {S.~J.}\ \bibnamefont {Tracy}},
  \bibinfo {author} {\bibfnamefont {D.~L.}\ \bibnamefont {Abernathy}}, \bibinfo
  {author} {\bibfnamefont {B.}~\bibnamefont {Fultz}}, \bibinfo {author}
  {\bibfnamefont {C.~W.}\ \bibnamefont {Li}}, \bibinfo {author} {\bibfnamefont
  {X.}~\bibnamefont {Tang}}, \ and\ \bibinfo {author} {\bibfnamefont {J.~A.}\
  \bibnamefont {Mun}},\ }\href {\doibase 10.1103/PhysRevLett.107.195504}
  {\bibfield  {journal} {\bibinfo  {journal} {Physical Review Letters}\
  }\textbf {\bibinfo {volume} {107}},\ \bibinfo {pages} {195504} (\bibinfo
  {year} {2011})}\BibitemShut {NoStop}%
\bibitem [{\citenamefont {Schick}\ and\ \citenamefont
  {Rappe}(2016)}]{Schick2016}%
  \BibitemOpen
  \bibfield  {author} {\bibinfo {author} {\bibfnamefont {J.~T.}\ \bibnamefont
  {Schick}}\ and\ \bibinfo {author} {\bibfnamefont {A.~M.}\ \bibnamefont
  {Rappe}},\ }\href {\doibase 10.1103/PhysRevB.93.214304} {\bibfield  {journal}
  {\bibinfo  {journal} {Physical Review B}\ }\textbf {\bibinfo {volume} {93}},\
  \bibinfo {pages} {1} (\bibinfo {year} {2016})},\ \Eprint
  {http://arxiv.org/abs/1604.07785} {arXiv:1604.07785} \BibitemShut {NoStop}%
\bibitem [{\citenamefont {Oba}\ \emph {et~al.}(2018)\citenamefont {Oba},
  \citenamefont {Tadano}, \citenamefont {Akashi},\ and\ \citenamefont
  {Tsuneyuki}}]{Oba2018}%
  \BibitemOpen
  \bibfield  {author} {\bibinfo {author} {\bibfnamefont {Y.}~\bibnamefont
  {Oba}}, \bibinfo {author} {\bibfnamefont {T.}~\bibnamefont {Tadano}},
  \bibinfo {author} {\bibfnamefont {R.}~\bibnamefont {Akashi}}, \ and\ \bibinfo
  {author} {\bibfnamefont {S.}~\bibnamefont {Tsuneyuki}},\ }\href {\doibase
  10.1103/PhysRevMaterials.3.033601} {\ \textbf {\bibinfo {volume} {033601}},\
  \bibinfo {pages} {1} (\bibinfo {year} {2018})},\ \Eprint
  {http://arxiv.org/abs/1810.08800} {arXiv:1810.08800} \BibitemShut {NoStop}%
\bibitem [{\citenamefont {Sanson}(2018)}]{Sanson2018}%
  \BibitemOpen
  \bibfield  {author} {\bibinfo {author} {\bibfnamefont {A.}~\bibnamefont
  {Sanson}},\ }\href {\doibase 10.1080/21663831.2019.1621957} {\bibfield
  {journal} {\bibinfo  {journal} {Materials Research Letters}\ }\textbf
  {\bibinfo {volume} {3831}},\ \bibinfo {pages} {412} (\bibinfo {year}
  {2018})},\ \Eprint {http://arxiv.org/abs/1809.04499} {arXiv:1809.04499}
  \BibitemShut {NoStop}%
\bibitem [{\citenamefont {Glazer}(1972)}]{Glazer1972}%
  \BibitemOpen
  \bibfield  {author} {\bibinfo {author} {\bibfnamefont {A.~M.}\ \bibnamefont
  {Glazer}},\ }\href {\doibase 10.1107/S0567740872007976} {\bibfield  {journal}
  {\bibinfo  {journal} {Acta Crystallographica Section B Structural
  Crystallography and Crystal Chemistry}\ }\textbf {\bibinfo {volume} {28}},\
  \bibinfo {pages} {3384} (\bibinfo {year} {1972})}\BibitemShut {NoStop}%
\bibitem [{\citenamefont {Senn}\ \emph {et~al.}(2016)\citenamefont {Senn},
  \citenamefont {Keen}, \citenamefont {Lucas}, \citenamefont {Hriljac},\ and\
  \citenamefont {Goodwin}}]{Senn2016}%
  \BibitemOpen
  \bibfield  {author} {\bibinfo {author} {\bibfnamefont {M.~S.}\ \bibnamefont
  {Senn}}, \bibinfo {author} {\bibfnamefont {D.~A.}\ \bibnamefont {Keen}},
  \bibinfo {author} {\bibfnamefont {T.~C.}\ \bibnamefont {Lucas}}, \bibinfo
  {author} {\bibfnamefont {J.~A.}\ \bibnamefont {Hriljac}}, \ and\ \bibinfo
  {author} {\bibfnamefont {A.~L.}\ \bibnamefont {Goodwin}},\ }\href {\doibase
  10.1103/PhysRevLett.116.207602} {\bibfield  {journal} {\bibinfo  {journal}
  {Physical Review Letters}\ }\textbf {\bibinfo {volume} {116}},\ \bibinfo
  {pages} {1} (\bibinfo {year} {2016})},\ \Eprint
  {http://arxiv.org/abs/1512.03643} {arXiv:1512.03643} \BibitemShut {NoStop}%
\bibitem [{\citenamefont {Dippel}\ \emph {et~al.}(2015)\citenamefont {Dippel},
  \citenamefont {Liermann}, \citenamefont {Delitz}, \citenamefont {Walter},
  \citenamefont {Schulte-Schrepping}, \citenamefont {Seeck},\ and\
  \citenamefont {Franz}}]{Dippel2015}%
  \BibitemOpen
  \bibfield  {author} {\bibinfo {author} {\bibfnamefont {A.-C.}\ \bibnamefont
  {Dippel}}, \bibinfo {author} {\bibfnamefont {H.-P.}\ \bibnamefont
  {Liermann}}, \bibinfo {author} {\bibfnamefont {J.~T.}\ \bibnamefont
  {Delitz}}, \bibinfo {author} {\bibfnamefont {P.}~\bibnamefont {Walter}},
  \bibinfo {author} {\bibfnamefont {H.}~\bibnamefont {Schulte-Schrepping}},
  \bibinfo {author} {\bibfnamefont {O.~H.}\ \bibnamefont {Seeck}}, \ and\
  \bibinfo {author} {\bibfnamefont {H.}~\bibnamefont {Franz}},\ }\href
  {\doibase 10.1107/s1600577515002222} {\bibfield  {journal} {\bibinfo
  {journal} {Journal of Synchrotron Radiation}\ }\textbf {\bibinfo {volume}
  {22}},\ \bibinfo {pages} {675} (\bibinfo {year} {2015})}\BibitemShut
  {NoStop}%
\bibitem [{\citenamefont {Basham}\ \emph {et~al.}(2015)\citenamefont {Basham},
  \citenamefont {Filik}, \citenamefont {Wharmby}, \citenamefont {Chang},
  \citenamefont {{El Kassaby}}, \citenamefont {Gerring}, \citenamefont
  {Aishima}, \citenamefont {Levik}, \citenamefont {Pulford}, \citenamefont
  {Sikharulidze}, \citenamefont {Sneddon}, \citenamefont {Webber},
  \citenamefont {Dhesi}, \citenamefont {Maccherozzi}, \citenamefont {Svensson},
  \citenamefont {Brockhauser}, \citenamefont {N{\'{a}}ray},\ and\ \citenamefont
  {Ashton}}]{Basham2015}%
  \BibitemOpen
  \bibfield  {author} {\bibinfo {author} {\bibfnamefont {M.}~\bibnamefont
  {Basham}}, \bibinfo {author} {\bibfnamefont {J.}~\bibnamefont {Filik}},
  \bibinfo {author} {\bibfnamefont {M.~T.}\ \bibnamefont {Wharmby}}, \bibinfo
  {author} {\bibfnamefont {P.~C.}\ \bibnamefont {Chang}}, \bibinfo {author}
  {\bibfnamefont {B.}~\bibnamefont {{El Kassaby}}}, \bibinfo {author}
  {\bibfnamefont {M.}~\bibnamefont {Gerring}}, \bibinfo {author} {\bibfnamefont
  {J.}~\bibnamefont {Aishima}}, \bibinfo {author} {\bibfnamefont
  {K.}~\bibnamefont {Levik}}, \bibinfo {author} {\bibfnamefont {B.~C.}\
  \bibnamefont {Pulford}}, \bibinfo {author} {\bibfnamefont {I.}~\bibnamefont
  {Sikharulidze}}, \bibinfo {author} {\bibfnamefont {D.}~\bibnamefont
  {Sneddon}}, \bibinfo {author} {\bibfnamefont {M.}~\bibnamefont {Webber}},
  \bibinfo {author} {\bibfnamefont {S.~S.}\ \bibnamefont {Dhesi}}, \bibinfo
  {author} {\bibfnamefont {F.}~\bibnamefont {Maccherozzi}}, \bibinfo {author}
  {\bibfnamefont {O.}~\bibnamefont {Svensson}}, \bibinfo {author}
  {\bibfnamefont {S.}~\bibnamefont {Brockhauser}}, \bibinfo {author}
  {\bibfnamefont {G.}~\bibnamefont {N{\'{a}}ray}}, \ and\ \bibinfo {author}
  {\bibfnamefont {A.~W.}\ \bibnamefont {Ashton}},\ }\href {\doibase
  10.1107/S1600577515002283} {\bibfield  {journal} {\bibinfo  {journal}
  {Journal of Synchrotron Radiation}\ }\textbf {\bibinfo {volume} {22}},\
  \bibinfo {pages} {853} (\bibinfo {year} {2015})}\BibitemShut {NoStop}%
\bibitem [{\citenamefont {McLain}\ \emph {et~al.}(2012)\citenamefont {McLain},
  \citenamefont {Bowron}, \citenamefont {Hannon},\ and\ \citenamefont
  {Soper}}]{McLain2012}%
  \BibitemOpen
  \bibfield  {author} {\bibinfo {author} {\bibfnamefont {S.~E.}\ \bibnamefont
  {McLain}}, \bibinfo {author} {\bibfnamefont {D.~T.}\ \bibnamefont {Bowron}},
  \bibinfo {author} {\bibfnamefont {A.~C.}\ \bibnamefont {Hannon}}, \ and\
  \bibinfo {author} {\bibfnamefont {A.~K.}\ \bibnamefont {Soper}},\ }\href@noop
  {} {\  (\bibinfo {year} {2012})}\BibitemShut {NoStop}%
\bibitem [{\citenamefont {Qiu}\ \emph {et~al.}(2004)\citenamefont {Qiu},
  \citenamefont {Thompson},\ and\ \citenamefont {Billinge}}]{Qiu2004}%
  \BibitemOpen
  \bibfield  {author} {\bibinfo {author} {\bibfnamefont {X.}~\bibnamefont
  {Qiu}}, \bibinfo {author} {\bibfnamefont {J.~W.}\ \bibnamefont {Thompson}}, \
  and\ \bibinfo {author} {\bibfnamefont {S.~J.}\ \bibnamefont {Billinge}},\
  }\href {\doibase 10.1107/S0021889804011744} {\bibfield  {journal} {\bibinfo
  {journal} {Journal of Applied Crystallography}\ }\textbf {\bibinfo {volume}
  {37}},\ \bibinfo {pages} {678} (\bibinfo {year} {2004})}\BibitemShut
  {NoStop}%
\bibitem [{\citenamefont {Popuri}\ \emph {et~al.}(2019)\citenamefont {Popuri},
  \citenamefont {Decourt}, \citenamefont {McNulty}, \citenamefont {Pollet},
  \citenamefont {Fortes}, \citenamefont {Morrison}, \citenamefont {Senn},\ and\
  \citenamefont {Bos}}]{Popuri2019}%
  \BibitemOpen
  \bibfield  {author} {\bibinfo {author} {\bibfnamefont {S.~R.}\ \bibnamefont
  {Popuri}}, \bibinfo {author} {\bibfnamefont {R.}~\bibnamefont {Decourt}},
  \bibinfo {author} {\bibfnamefont {J.~A.}\ \bibnamefont {McNulty}}, \bibinfo
  {author} {\bibfnamefont {M.}~\bibnamefont {Pollet}}, \bibinfo {author}
  {\bibfnamefont {A.~D.}\ \bibnamefont {Fortes}}, \bibinfo {author}
  {\bibfnamefont {F.~D.}\ \bibnamefont {Morrison}}, \bibinfo {author}
  {\bibfnamefont {M.~S.}\ \bibnamefont {Senn}}, \ and\ \bibinfo {author}
  {\bibfnamefont {J.~W.}\ \bibnamefont {Bos}},\ }\href {\doibase
  10.1021/acs.jpcc.8b10520} {\bibfield  {journal} {\bibinfo  {journal} {Journal
  of Physical Chemistry C}\ }\textbf {\bibinfo {volume} {123}},\ \bibinfo
  {pages} {5198} (\bibinfo {year} {2019})}\BibitemShut {NoStop}%
\bibitem [{\citenamefont {Campbell}\ \emph {et~al.}(2006)\citenamefont
  {Campbell}, \citenamefont {Stokes}, \citenamefont {Tanner},\ and\
  \citenamefont {Hatch}}]{Campbell2006}%
  \BibitemOpen
  \bibfield  {author} {\bibinfo {author} {\bibfnamefont {B.~J.}\ \bibnamefont
  {Campbell}}, \bibinfo {author} {\bibfnamefont {H.~T.}\ \bibnamefont
  {Stokes}}, \bibinfo {author} {\bibfnamefont {D.~E.}\ \bibnamefont {Tanner}},
  \ and\ \bibinfo {author} {\bibfnamefont {D.~M.}\ \bibnamefont {Hatch}},\
  }\href {\doibase 10.1107/S0021889806014075} {\bibfield  {journal} {\bibinfo
  {journal} {Journal of Applied Crystallography}\ }\textbf {\bibinfo {volume}
  {39}},\ \bibinfo {pages} {607} (\bibinfo {year} {2006})}\BibitemShut
  {NoStop}%
\bibitem [{\citenamefont {Coelho}\ \emph {et~al.}(2015)\citenamefont {Coelho},
  \citenamefont {Chater},\ and\ \citenamefont {Kern}}]{Coelho2015}%
  \BibitemOpen
  \bibfield  {author} {\bibinfo {author} {\bibfnamefont {A.~A.}\ \bibnamefont
  {Coelho}}, \bibinfo {author} {\bibfnamefont {P.~A.}\ \bibnamefont {Chater}},
  \ and\ \bibinfo {author} {\bibfnamefont {A.}~\bibnamefont {Kern}},\ }\href
  {\doibase 10.1107/s1600576715007487} {\bibfield  {journal} {\bibinfo
  {journal} {Journal of Applied Crystallography}\ }\textbf {\bibinfo {volume}
  {48}},\ \bibinfo {pages} {869} (\bibinfo {year} {2015})}\BibitemShut
  {NoStop}%
\bibitem [{\citenamefont {Momma}\ and\ \citenamefont
  {Izumi}(2011)}]{Momma2011}%
  \BibitemOpen
  \bibfield  {author} {\bibinfo {author} {\bibfnamefont {K.}~\bibnamefont
  {Momma}}\ and\ \bibinfo {author} {\bibfnamefont {F.}~\bibnamefont {Izumi}},\
  }\href@noop {} {\bibfield  {journal} {\bibinfo  {journal} {Journal of Applied
  Crystallography}\ }\textbf {\bibinfo {volume} {44}},\ \bibinfo {pages} {1272}
  (\bibinfo {year} {2011})}\BibitemShut {NoStop}%
\bibitem [{\citenamefont {Farrow}\ \emph {et~al.}(2007)\citenamefont {Farrow},
  \citenamefont {Juhas}, \citenamefont {Liu}, \citenamefont {Bryndin},
  \citenamefont {Boin}, \citenamefont {Bloch}, \citenamefont {Proffen},\ and\
  \citenamefont {Billinge}}]{Farrow2007}%
  \BibitemOpen
  \bibfield  {author} {\bibinfo {author} {\bibfnamefont {C.~L.}\ \bibnamefont
  {Farrow}}, \bibinfo {author} {\bibfnamefont {P.}~\bibnamefont {Juhas}},
  \bibinfo {author} {\bibfnamefont {J.~W.}\ \bibnamefont {Liu}}, \bibinfo
  {author} {\bibfnamefont {D.}~\bibnamefont {Bryndin}}, \bibinfo {author}
  {\bibfnamefont {E.~S.}\ \bibnamefont {Boin}}, \bibinfo {author}
  {\bibfnamefont {J.}~\bibnamefont {Bloch}}, \bibinfo {author} {\bibfnamefont
  {T.}~\bibnamefont {Proffen}}, \ and\ \bibinfo {author} {\bibfnamefont
  {S.~J.}\ \bibnamefont {Billinge}},\ }\href {\doibase
  10.1088/0953-8984/19/33/335219} {\bibfield  {journal} {\bibinfo  {journal}
  {Journal of Physics Condensed Matter}\ }\textbf {\bibinfo {volume} {19}},\
  \bibinfo {pages} {1} (\bibinfo {year} {2007})}\BibitemShut {NoStop}%
\bibitem [{\citenamefont {Hu}\ \emph {et~al.}(2016{\natexlab{c}})\citenamefont
  {Hu}, \citenamefont {Chen}, \citenamefont {Xu}, \citenamefont {Wang},
  \citenamefont {Han}, \citenamefont {Ren}, \citenamefont {Pan}, \citenamefont
  {Rong}, \citenamefont {Huang}, \citenamefont {Deng}, \citenamefont {Li},\
  and\ \citenamefont {Xing}}]{Hu2016b}%
  \BibitemOpen
  \bibfield  {author} {\bibinfo {author} {\bibfnamefont {L.}~\bibnamefont
  {Hu}}, \bibinfo {author} {\bibfnamefont {J.}~\bibnamefont {Chen}}, \bibinfo
  {author} {\bibfnamefont {J.}~\bibnamefont {Xu}}, \bibinfo {author}
  {\bibfnamefont {N.}~\bibnamefont {Wang}}, \bibinfo {author} {\bibfnamefont
  {F.}~\bibnamefont {Han}}, \bibinfo {author} {\bibfnamefont {Y.}~\bibnamefont
  {Ren}}, \bibinfo {author} {\bibfnamefont {Z.}~\bibnamefont {Pan}}, \bibinfo
  {author} {\bibfnamefont {Y.}~\bibnamefont {Rong}}, \bibinfo {author}
  {\bibfnamefont {R.}~\bibnamefont {Huang}}, \bibinfo {author} {\bibfnamefont
  {J.}~\bibnamefont {Deng}}, \bibinfo {author} {\bibfnamefont {L.}~\bibnamefont
  {Li}}, \ and\ \bibinfo {author} {\bibfnamefont {X.}~\bibnamefont {Xing}},\
  }\href {\doibase 10.1021/jacs.6b08746} {\bibfield  {journal} {\bibinfo
  {journal} {Journal of the American Chemical Society}\ }\textbf {\bibinfo
  {volume} {138}},\ \bibinfo {pages} {14530} (\bibinfo {year}
  {2016}{\natexlab{c}})}\BibitemShut {NoStop}%
\bibitem [{\citenamefont {Yang}\ \emph {et~al.}(2016)\citenamefont {Yang},
  \citenamefont {Tong}, \citenamefont {Lin}, \citenamefont {Guo}, \citenamefont
  {Zhang}, \citenamefont {Wang}, \citenamefont {Wu}, \citenamefont {Lin},
  \citenamefont {Huang}, \citenamefont {Xu}, \citenamefont {Song},\ and\
  \citenamefont {Sun}}]{Yang2016}%
  \BibitemOpen
  \bibfield  {author} {\bibinfo {author} {\bibfnamefont {C.}~\bibnamefont
  {Yang}}, \bibinfo {author} {\bibfnamefont {P.}~\bibnamefont {Tong}}, \bibinfo
  {author} {\bibfnamefont {J.~C.}\ \bibnamefont {Lin}}, \bibinfo {author}
  {\bibfnamefont {X.~G.}\ \bibnamefont {Guo}}, \bibinfo {author} {\bibfnamefont
  {K.}~\bibnamefont {Zhang}}, \bibinfo {author} {\bibfnamefont
  {M.}~\bibnamefont {Wang}}, \bibinfo {author} {\bibfnamefont {Y.}~\bibnamefont
  {Wu}}, \bibinfo {author} {\bibfnamefont {S.}~\bibnamefont {Lin}}, \bibinfo
  {author} {\bibfnamefont {P.~C.}\ \bibnamefont {Huang}}, \bibinfo {author}
  {\bibfnamefont {W.}~\bibnamefont {Xu}}, \bibinfo {author} {\bibfnamefont
  {W.~H.}\ \bibnamefont {Song}}, \ and\ \bibinfo {author} {\bibfnamefont
  {Y.~P.}\ \bibnamefont {Sun}},\ }\href {\doibase 10.1063/1.4959083} {\bibfield
   {journal} {\bibinfo  {journal} {Applied Physics Letters}\ }\textbf {\bibinfo
  {volume} {109}},\ \bibinfo {pages} {3} (\bibinfo {year} {2016})}\BibitemShut
  {NoStop}%
\bibitem [{\citenamefont {Prisco}\ \emph {et~al.}(2013)\citenamefont {Prisco},
  \citenamefont {Romao}, \citenamefont {Rizzo}, \citenamefont {White},\ and\
  \citenamefont {Marinkovic}}]{Prisco2013}%
  \BibitemOpen
  \bibfield  {author} {\bibinfo {author} {\bibfnamefont {L.~P.}\ \bibnamefont
  {Prisco}}, \bibinfo {author} {\bibfnamefont {C.~P.}\ \bibnamefont {Romao}},
  \bibinfo {author} {\bibfnamefont {F.}~\bibnamefont {Rizzo}}, \bibinfo
  {author} {\bibfnamefont {M.~A.}\ \bibnamefont {White}}, \ and\ \bibinfo
  {author} {\bibfnamefont {B.~A.}\ \bibnamefont {Marinkovic}},\ }\href
  {\doibase 10.1007/s10853-012-7076-9} {\bibfield  {journal} {\bibinfo
  {journal} {Journal of Materials Science}\ }\textbf {\bibinfo {volume} {48}},\
  \bibinfo {pages} {2986} (\bibinfo {year} {2013})}\BibitemShut {NoStop}%
\bibitem [{\citenamefont {Dove}\ \emph {et~al.}(2019)\citenamefont {Dove},
  \citenamefont {Du}, \citenamefont {Keen}, \citenamefont {Tucker},\ and\
  \citenamefont {Phillips}}]{Dove2019}%
  \BibitemOpen
  \bibfield  {author} {\bibinfo {author} {\bibfnamefont {M.~T.}\ \bibnamefont
  {Dove}}, \bibinfo {author} {\bibfnamefont {J.}~\bibnamefont {Du}}, \bibinfo
  {author} {\bibfnamefont {D.~A.}\ \bibnamefont {Keen}}, \bibinfo {author}
  {\bibfnamefont {M.~G.}\ \bibnamefont {Tucker}}, \ and\ \bibinfo {author}
  {\bibfnamefont {A.~E.}\ \bibnamefont {Phillips}},\ }\href
  {http://arxiv.org/abs/1905.09250} {\  (\bibinfo {year} {2019})},\ \Eprint
  {http://arxiv.org/abs/1905.09250} {arXiv:1905.09250} \BibitemShut {NoStop}%
\bibitem [{\citenamefont {Wendt}\ \emph {et~al.}(2019)\citenamefont {Wendt},
  \citenamefont {Bozin}, \citenamefont {Neuefeind}, \citenamefont {Page},
  \citenamefont {Ku}, \citenamefont {Wang}, \citenamefont {Fultz},
  \citenamefont {Tkachenko},\ and\ \citenamefont {Zaliznyak}}]{Wendt2019}%
  \BibitemOpen
  \bibfield  {author} {\bibinfo {author} {\bibfnamefont {D.}~\bibnamefont
  {Wendt}}, \bibinfo {author} {\bibfnamefont {E.}~\bibnamefont {Bozin}},
  \bibinfo {author} {\bibfnamefont {J.}~\bibnamefont {Neuefeind}}, \bibinfo
  {author} {\bibfnamefont {K.}~\bibnamefont {Page}}, \bibinfo {author}
  {\bibfnamefont {W.}~\bibnamefont {Ku}}, \bibinfo {author} {\bibfnamefont
  {L.}~\bibnamefont {Wang}}, \bibinfo {author} {\bibfnamefont {B.}~\bibnamefont
  {Fultz}}, \bibinfo {author} {\bibfnamefont {A.}~\bibnamefont {Tkachenko}}, \
  and\ \bibinfo {author} {\bibfnamefont {I.}~\bibnamefont {Zaliznyak}},\ }\href
  {http://arxiv.org/abs/1906.05213} {\ ,\ \bibinfo {pages} {1} (\bibinfo {year}
  {2019})},\ \Eprint {http://arxiv.org/abs/1906.05213} {arXiv:1906.05213}
  \BibitemShut {NoStop}%
\end{thebibliography}%
\end{document}